\documentclass[conference]{IEEEtran}
\IEEEoverridecommandlockouts
\usepackage{cite}
\usepackage{amsmath,amssymb,amsfonts}
\usepackage{algorithmic}
\usepackage{graphicx}
\usepackage{textcomp}
\usepackage{xcolor}
\usepackage{cite}
\usepackage{import}
\usepackage{subcaption}
\usepackage{stmaryrd}
\usepackage{booktabs}
\usepackage{multirow}
\usepackage[table,xcdraw]{xcolor}
\usepackage{makecell}
\usepackage{float}
\usepackage{lipsum}
\usepackage{mathtools}
\usepackage{flushend}

\def\BibTeX{{\rm B\kern-.05em{\sc i\kern-.025em b}\kern-.08em
    T\kern-.1667em\lower.7ex\hbox{E}\kern-.125emX}}
\begin{document}

\title{Gaussian Phase Noise Effects on Hybrid Precoding MIMO Systems for Sub-THz Transmission}
\author{
\IEEEauthorblockN{
Yaya Bello\textsuperscript{1},
Yahia Medjahdi\textsuperscript{1},
Laurent Clavier\textsuperscript{1,2},
Arthur Louchart\textsuperscript{1},
}
\IEEEauthorblockA{
\textsuperscript{1}IMT Nord Europe, Institut Mines T\'el\'ecom, Center for Digital Systems, F-59653 Villeneuve d’Ascq, France\\
}
\IEEEauthorblockA{
\textsuperscript{2}Inria, Villeneuve d'Ascq, France\\
Email: yaya.bello@imt-nord-europe.fr
}
}

\maketitle

\bstctlcite{IEEEexample:BSTcontrol}
\maketitle
\begin{abstract}
The sub-THz spectrum offers numerous advantages, including massive multiple-input multiple-output~(MIMO) technology with large antenna arrays that enhance spectral efficiency (SE) of future systems. Hybrid precoding~(HP) thus emerges as a cost-effective alternative to fully digital precoding regarding complexity and energy consumption. However, sub-THz frequencies introduce hardware challenges, particularly phase noise~(PN) from local oscillators (LOs). We analyze~PN impact on~MIMO systems using~HP, leveraging singular value decomposition and common LO architecture. We adopt the Gaussian PN~(GPN) model, recognized as accurate for describing~PN behavior in sub-THz transmissions. We derive a lower bound on achievable SE and provide closed-form bit error rate expressions for quadrature amplitude modulation (QAM)—specifically 4-QAM and~16-QAM—under high-SNR and strong GPN conditions. These analytical results are validated through Monte Carlo simulations. We show that~GPN can be effectively counteracted with a single pilot symbol in single-user MIMO systems, unlike single-input single-output systems where mitigation proves infeasible. Simulation results compare conventional~QAM against \mbox{polar-QAM} tailored for GPN-impaired systems. Finally, we introduce perspectives for further improvements in performance and energy efficiency.  

\end{abstract}

\section{Introduction}
\IEEEPARstart{T}{o} meet the surging demand for high-speed multimedia, wireless capacities must increase exponentially~\cite{Andrews}. Future architectures target an approximate thousand-fold increase w.r.t. 5G improvements~\cite{Tong}. Key strategies include: (i) boosting physical~(PHY) layer spectral efficiency~(SE) via massive multiple-input multiple-output (MIMO)~\cite{Rong}, and (ii) expanding bandwidth. Spectrum scarcity in contemporary cellular infrastructures constitutes a critical limiting factor that fundamentally constrains any prospective enhancement in network capacity. Frequency bands in the sub‑terahertz (sub‑THz) range, specifically spanning from $90$~GHz to $300$~GHz, have recently emerged as leading candidates for additional spectrum resources for the next generation of cellular systems~\cite{Rappaport}. These bands offer the prospect of exceptionally large contiguous bandwidths and thus enabling unprecedented data throughput capabilities~\cite{Dore2018}. However, transmitting in this frequency band presents several obstacles, such as huge channel path loss, high sampling frequency required from digital-to-analog~(DAC) and analog-to-digital~(ADC) converters, and the phase noise~(PN) generated by the local oscillator~(LO) at both the transmitter~(Tx) and receiver~(Rx)~\cite{adc_survey}. Owing to the intrinsically short wavelength of millimeter‑wave (mmWave) signals, sub-THz MIMO precoding could exploit very large‑scale antenna arrays at the Tx and~Rx to achieve substantial beamforming gains. These gains effectively mitigate severe propagation losses while enabling the synthesis of highly directional transmission beams~\cite{Pometcu,Pometcu2,Xing2}.

In conventional MIMO architectures, precoding operations are predominantly executed digitally within the baseband, enabling precise control over both amplitude and phase components of transmitted signals~\cite{Hoydis}. Nevertheless, implementing fully digital precoding~(FDP) necessitates the deployment of a radio frequency (RF) chain for each antenna element, encompassing components such as mixers and ADC/DAC converters. Although the diminutive wavelengths characteristic of sub-THz bands facilitate the integration of very large antenna arrays, the substantial cost and power consumption associated with equipping each element with a dedicated RF chain renders~FDP impractical~\cite{Busari,Kebede}. To address these constraints inherent in mmWave MIMO systems, hybrid precoding~(HP) frameworks have attracted considerable research interest~\cite{Sun,Gao,Rappaport2}. Such architectures strategically combine a reduced number of RF chains interfacing a low-dimensional digital precoder with a high-dimensional analog precoder, thereby enabling efficient utilization of hardware resources while achieving near-optimal precoding performance.

PN originates from rapid, stochastic fluctuations in the phase of an oscillator signal induced by intrinsic device components. These short-term random phase variations disrupt the temporal stability of the waveform, leading to performance degradation~\cite{Demir}. Furthermore, PN can manifest as synchronization challenges in digitally clocked and sampled-data systems. The severity of PN escalates with increasing carrier frequency, thereby posing significant detriments to system reliability and efficiency, particularly in transmissions at sub-THz frequency ranges. Fundamentally, PN arises as the composite effect of multiple random processes. Some exhibiting temporal correlation, such as the Wiener process, and others uncorrelated, exemplified by additive white Gaussian noise. In the literature, models incorporating correlated behavior, including the Wiener PN model and the standardized 3GPP PN framework, have been extensively adopted~\cite{Levanen,Afshang,Tervo2}. The correlated component of PN can, in principle, be tracked and subsequently compensated to mitigate its degradation of radio link performance~\cite{Bello}. Nevertheless, as large signal bandwidths are feasible within sub‑THz frequency regimes, the influence of the residual uncorrelated Gaussian PN~(GPN) component becomes predominant, thus posing a more critical limitation on system performance~\cite{Bicais2}. 

In MIMO systems, two LO architectures are possible: (i) common LO~(CLO), where all RF chains share a centralized LO, and (ii) independent LO~(ILO), where each RF chain uses its own LO. As reported in~\cite{Thomas, Emil, Pitarokoilis}, the ILO architecture is less impacted by~PN than the CLO configuration. Consequently, we investigate in this work the GPN effect on HP single-user MIMO~(SU-MIMO) systems in terms of achievable rate and derive closed-form bite error rate~(BER) expressions in high SNR and strong GPN regimes focusing on CLO architecture. 

\subsection{Related Works}
The PN impact in HP MIMO systems has been investigated using both Wiener PN models~\cite{Zhang,Faragallah,Nguyen} and~GPN models~\cite{Corvaja,Corvaja2}. In~\cite{Zhang}, the authors highlighted the PN effect in multi-user MIMO systems employing zero-forcing~(ZF) precoding for downlink transmission under a~CLO architecture. They derived a lower bound on the achievable rate based on the channel covariance matrices used for channel estimation during the uplink transmission. However, utilizing the~ZF precoder requires the number of~RF chains to match the number of users, which, from an energy consumption perspective, degrades energy efficiency~(EE). In~\cite{Faragallah}, the authors proposed several PN estimation techniques for PN compensation, with the aim of enhancing the performance of a~SU-MIMO system operating with~ILO architecture. They assumed perfect channel state information~(CSI) knowledge, and HP matrices were designed following the approach presented in~\cite{Yu}. In~\cite{Corvaja}, the authors demonstrated the performance degradation of HP SU-MIMO systems employing singular value decomposition~(SVD) precoding in the presence of GPN. In~\cite{Corvaja2}, they further examined the impact of channel estimation errors and GPN on system performance, assuming fixed analog beamforming matrices to reduce design complexity. Unfortunately, no theoretical expressions of the achievable rate or~BER are derived in~\cite{Corvaja,Corvaja2}. Although~GPN is considered the most suitable model for describing~PN effects in sub-THz transmissions~\cite{Bicais2}, relatively few papers in the literature address its impact on~HP MIMO systems. To the best of our knowledge, no analytical BER expression has been derived for SVD-based HP SU-MIMO systems affected by GPN. This work therefore contributes such an analysis.

\begin{figure}[htpb]
   \centering      
    \includegraphics[width=1\columnwidth]{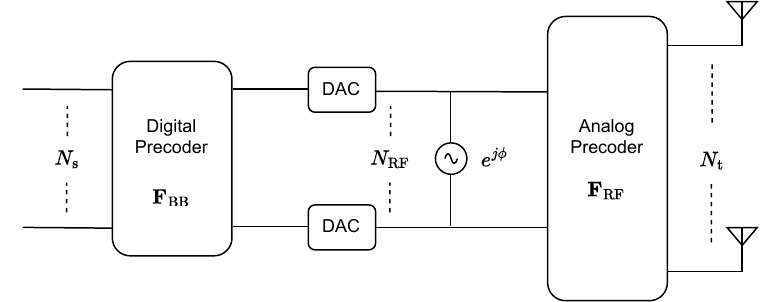}
    \caption{HP massive MIMO Tx system model considering CLO architecture.}
    \label{fig:system_model_clo_Tx}
    \vspace{-0.3cm}
\end{figure}

\subsection{Contributions}
The paper highlights the following contributions: 
\begin{itemize}
    \item Derive and analyze the theoretical expression of the lower bound on the achievable SE of HP SU-MIMO in the presence of GPN.
    \item Provide semi-analytical BER expressions of HP SU-MIMO in the presence of GPN for both quadrature amplitude modulation~(QAM) especially 4-QAM and 16-QAM, and polar-QAM~(PQAM) constellation schemes.
    \item Propose closed-form expressions of the average BER in high-SNR and strong GPN regimes and validate them via Monte Carlo simulations.
    \item Formulate semi-analytical expressions of the sum rate and of the~BER when PN is alleviated.
    \item Investigate and compare the system performance between conventional QAM modulation scheme and PQAM constellation, both with and without PN cancellation.
\end{itemize}

\subsection{Organization}
The remainder of the paper is as follows. Sec.~\ref{sec:sys_model} and Sec.~\ref{sec:pn_met} present the system model and the various closed-form expressions of the achievable~SE and BER, respectively. The simulation results are carried out in Sec.~\ref{sec:results}, discussions and perspectives are figured out in Sec.~\ref{sec:discussions}. We end up with the conclusion in Sec.~\ref{sec:conclusion}.

\subsection{Notations}
Higher boldface letters $\mathbf{A}$ denote matrices and lower boldface letters~$\mathbf{a}$ indicate column vectors where the scalar~$a_k$ is its $k^{th}$ element. The term~$a_{k_\rho}$ (resp.~$a_{k_\theta}$) denotes the magnitude value (resp. phase value) of the scalar~$a_k$. The term $\mathbb{E}\left\lbrace\cdot\right\rbrace$ denotes the expectation operator. The operators $\left \lvert\cdot \right \lvert$, $\text{arg}\left\lbrace\cdot \right\rbrace$, $\mathfrak{Re}\{\cdot\}$, $\mathfrak{Im}\{\cdot\}$, $\left (\cdot \right)^T$, $\left (\cdot \right)^{*}$, $(\cdot)^H$, $\Arrowvert \cdot \Arrowvert$ and $\Arrowvert \cdot \Arrowvert_F$ return respectively the magnitude, the phase, the real part, the imaginary part, the transpose, the conjugate, the Hermitian, the Euclidean norm and the Frobenius norm value of the argument. A Gaussian (resp. a complex Gaussian) random variable $d$ is denoted \mbox{$d \backsim \mathcal{N}\left(m,q \right)$} (resp.~$d \backsim \mathbb{C}\mathcal{N}\left(m,q \right)$), where $m$ is the mean and $q$ is the variance. A circular symmetric complex Gaussian random vector $\mathbf{d}$ is denoted $\mathbf{d} \backsim \mathbb{C}\mathcal{N}\left(\mathbf{m}, \mathbf{T} \right)$, where $\mathbf{m}$ is the mean and $\mathbf{T}$ is the covariance matrix. The term $\mathbf{I}_X$ stands for the identity matrix of size~$X$. Finally, $j \triangleq \sqrt{-1}$ defines the unit imaginary number.

\section{System Model} \label{sec:sys_model}
\subsection{Channel and PN Models} \label{subsec:chan}
The propagation environment in sub-THz frequencies is expected to be typically sparse thanks to the use of high gain and directive antennas focusing the energy in the desired direction and thereby, reduce multipath. We consider the Saleh-Valenzuela model which is the most commonly used channel model in mmWave transmissions~\cite{Ayach,Sohrabi,Hu,Li}~due to very high free-space path loss. Therefore, by assuming a half-wave spaced uniform linear array at both the Tx and Rx, the channel matrix $\mathbf{H}$ of size $N_{\text{r}} \times N_{\text{t}}$ is defined by~\cite{Lin}
\begin{equation}
    \mathbf{H} = \sqrt{\frac{N_\text{t} N_\text{r}}{N_\text{c} N_\text{R}}} \sum_{i=1}^{N_\text{c}} \sum_{l=1}^{N_\text{R}} \xi_{il}\mathbf{a}_{\text{r}}\left(\theta^{\text{r}}_{il}\right)\mathbf{a}_{\text{t}}\left(\theta^{\text{t}}_{il}\right)^H,
\end{equation}
where $N_\text{r}$, $N_\text{t}$, $N_\text{c}$ and $N_\text{R}$ are the number of receive antennas, the number of transmit antennas, the number of clusters and the number of rays within each cluster, respectively. The term $\xi_{il} \sim \mathcal{C}\mathcal{N}\left(0,1\right)$ denotes the complex channel gain of the $l^{th}$ ray in the $i^{th}$ propagation cluster. The vectors $\mathbf{a}_{\text{r}}\left(\theta^{\text{r}}_{il}\right)$ and $\mathbf{a}_{\text{t}}\left(\theta^{\text{t}}_{il}\right)$ represent the normalized responses of the transmit and receive antenna arrays, respectively, which are defined by 
\begin{equation}
\begin{aligned}
    \mathbf{a}_{\text{r}}\left(\theta^{\text{r}}_{il}\right) = \frac{1}{\sqrt{N_{\text{r}}}} \left[1 \; e^{j\pi\sin{\theta^{\text{r}}_{il}}}\; \cdots \; e^{j\pi\left(N_{\text{r}}-1\right)\sin{\theta^{\text{r}}_{il}}}\right]^T, \\
    \mathbf{a}_{\text{t}}\left(\theta^{\text{t}}_{il}\right) = \frac{1}{\sqrt{N_{\text{t}}}} \left[1 \; e^{j\pi\sin{\theta^{\text{t}}_{il}}}\; \cdots \; e^{j\pi\left(N_{\text{t}}-1\right)\sin{\theta^{\text{t}}_{il}}}\right]^T,
    \end{aligned}
\end{equation}
where $\theta^{\text{r}}_{il}$ and $\theta^{\text{t}}_{il}$ stand for the angle of arrival~(AoA) and departure~(AoD), respectively.
 
As long as the PN increases with the carrier frequency, the influence of the uncorrelated~GPN component becomes dominant with the increase of the signal bandwidth. Thus, we consider a GPN model where the $k^{th}$ PN sample generated at the $i^{th}$ RF chain is modeled by
\begin{equation}
\phi_i[k] \sim \mathcal{N}\left(0,\sigma^2_{\phi}\right) \quad \text{and} \quad \varphi_i[k] \sim \mathcal{N}\left(0,\sigma^2_{\varphi}\right),
    \label{}
\end{equation}
where $\sigma^2_{\phi}$ and $\sigma^2_{\varphi}$ denote the GPN variance at the~Tx and~Rx, respectively. Three GPN levels are considered: strong GPN for $\sigma^2_\psi = 10^{-1}$, medium GPN for $\sigma^2_\psi = 10^{-2}$ and low GPN for $\sigma^2_\psi = 10^{-3}$ assuming~$\sigma^2_\psi=\sigma^2_\phi + \sigma^2_\varphi$ with $\sigma^2_\phi = \sigma^2_\varphi$. Finally, we adopt a coherent~Rx design,~\textit{i.e.}, with perfect synchronization in both time and frequency.
 
\subsection{HP SU-MIMO Systems with CLO Architecture} 
In this paper, we consider a SVD-based HP SU-MIMO system under CLO scheme where all the RF chains share the same LO as figured out in~Fig.~\ref{fig:system_model_clo_Tx}\footnote{Obviously, the Rx system model is the reverse of the processing steps performed at the Tx.}. Regarding analog precoding, the mapping configuration between RF chains and antenna elements determines the required number of phase shifters~(PSs). Based on this configuration,~HP transceiver architectures can be broadly classified into two categories: fully-connected~(FC) and partially-connected~(PC) topologies, as depicted in~\cite{Yu}. We consider the FC architecture where each RF chain interfaces with all antenna elements via a complete network of PSs, thereby enabling maximum beamforming gain and improved array directivity. For the analog precoding and combining design, we assume the low complexity algorithm named phase extraction-AltMin~(PE-AltMin)~\cite{Yu} achieving performance levels almost similar to those of~FDP systems in terms of~SE when the number of RF chain~$N_{\text{RF}}$ is equal to the number of data stream~$N_{\text{s}}$ as considered in this work. 

Sharing the same LO leads to a same PN process on any RF chain, \textit{i.e.},
\begin{equation}
    \forall i \in \left\lbrace 0,N_\text{RF}-1\right\rbrace, \quad \phi_i = \phi \quad \text{and} \quad \varphi_i = \varphi \cdot
\end{equation}
Therefore, the received signal expression is given by
\begin{equation}
    \begin{aligned}
        \mathbf{r}_{\text{PN}} &= e^{j\left(\phi+\varphi\right)} \mathbf{U}^H_{\text{BB}} \mathbf{W}^H_{\text{RF}} \mathbf{H} \mathbf{F}_{\text{RF}} \mathbf{F}_{\text{BB}}\mathbf{s} + e^{j\varphi} \mathbf{U}^H_{\text{BB}} \mathbf{W}^H_{\text{RF}}\mathbf{n}\\
        &= e^{j\psi} \mathbf{V} \mathbf{s} + e^{j\varphi} \mathbf{U}^H_{\text{BB}} \mathbf{W}^H_{\text{RF}}\mathbf{n},
    \end{aligned}
    \label{eq:rcvd_hybp_pn_clo}
\end{equation}
where~$\mathbf{V} = \mathbf{U}^H_{\text{BB}} \mathbf{W}^H_{\text{RF}} \mathbf{H} \mathbf{F}_{\text{RF}} \mathbf{F}_{\text{BB}}$. The terms $\mathbf{F}_{\text{BB}}$ and $\mathbf{F}_{\text{RF}}$ denote the digital precoder matrix of size $N_{\text{RF}} \times N_{\text{s}}$ and the analog precoder matrix of size $N_{\text{t}} \times N_{\text{RF}}$, respectively. The terms $\mathbf{U}_{\text{BB}}$ and $\mathbf{W}_{\text{RF}}$ are the digital decoder matrix of size $N_{\text{RF}} \times N_{\text{s}}$ and the analog decoder matrix of size $N_{\text{r}} \times N_{\text{RF}}$, respectively. The term $\mathbf{s}$ represents the transmitted signal such that $\mathbf{s} \sim \mathbb{C} \mathcal{N}\left(\mathbf{0},\mathbf{I}_{N_\text{s}}\right)$ and, $\mathbf{n} \sim \mathbb{C} \mathcal{N}\left(\mathbf{0},\sigma^2 \mathbf{I}_{N_\text{r}}\right)$ denotes the thermal noise which is independent and identically distributed. The term $\psi=\phi+\varphi \sim \mathcal{N}\left(0,\sigma^2_\psi = \sigma^2_\phi + \sigma^2_\varphi \right)$ stands for the sum of the PN processes.

\section{Derivation of bit error rate and Achievable Rate Expressions} \label{sec:pn_met}
In this section, we derive the theoretical lower bound expression on the achievable~SE. We also provide semi-analytical and closed-form BER expressions of SVD-based HP SU-MIMO systems with and without GPN.  

\subsection{HP Systems without PN}
Let us express the received signal at the~$k^{th}$ stream without PN by  
\begin{equation}
 \label{eq:the_rvd_signal_str}
    \begin{aligned}
        r_k &= \mathbf{u}_k^{\text{BB}^H} \mathbf{W}^H_{\text{RF}} \mathbf{H} \mathbf{F}_{\text{RF}} \mathbf{f}_k^{\text{BB}} s_k + \sum_{i=1,i\neq k}^{N_{\text{s}}} \mathbf{u}_k^{\text{BB}^H} \mathbf{W}^H_{\text{RF}} \mathbf{H} \mathbf{F}_{\text{RF}} \mathbf{f}_i^{\text{BB}} s_i \\ 
        &+ \mathbf{u}_k^{\text{BB}^H} \mathbf{W}^H_{\text{RF}} \mathbf{n}, 
    \end{aligned}
\end{equation}
where~$\mathbf{u}_k^{\text{BB}}$ and~$\mathbf{f}_k^{\text{BB}}$ denote the $k^{th}$ column of~$\mathbf{U}_{\text{BB}}$ and~$\mathbf{F}_{\text{BB}}$, respectively. The second term in~\eqref{eq:the_rvd_signal_str} represents the inter-stream interference~(ISI), \textit{i.e.}, the interference of the other streams on the stream~$k$. Despite the optimality of analog/digital precoders and decoders put forward in~\cite{Yu}, the BER performance of the HP system is limited since the optimization criterion does not take into account the error minimization. Several works emerged and introduced new algorithms for hybrid precoders and decoders regarding the BER minimization~\cite{Lin,Li}. Nevertheless, these algorithms bring more computational complexity. In this work, we consider, as suggested in~\cite{Corvaja,Corvaja2}, that the digital precoder and decoder are derived from the SVD of the equivalent channel~$\mathbf{H}_{\text{eq}}=\mathbf{W}_{\text{RF}}^H \mathbf{H} \mathbf{F}_{\text{RF}}=\mathbf{U}_{\text{BB}} \mathbf{V}\mathbf{F}^H_{\text{BB}}$. Thus, the matrix~$\mathbf{V}$ represents the diagonal matrix of singular values of the matrix~$\mathbf{H}_{\text{eq}}$ and therefore the ISI term~$\sum_{i=1,i\neq k}^{N_{\text{s}}} \mathbf{u}_k^{\text{BB}^H} \mathbf{H}_{\text{eq}} \mathbf{f}_i^{\text{BB}} s_i = 0 \cdot$

Additionally, since the objective is to achieve the same SE performance as in~FDP, the digital precoder $\mathbf{F}_{\text{BB}}$ is normalized by a factor $\rho = \frac{\sqrt{N_{\text{s}}}}{\Arrowvert \mathbf{F}_{\text{RF}} \mathbf{F}_{\text{BB}} \Arrowvert_F }$ in order to ensure that the signal power after the two precoding stages (digital and analog) remains equal to the signal power after the digital precoder~$\mathbf{F}_{\text{opt}}$. Thus, by posing $\mathbf{H}_{\text{eq}} = \mathbf{W}^H_{\text{RF}} \mathbf{H} \mathbf{F}_{\text{RF}}$, the expression~\eqref{eq:the_rvd_signal_str} becomes 
\begin{equation}
 \label{eq:the_rvd_signal_str2}
    \begin{aligned}
        r_k &= \rho \mathbf{u}_k^{\text{BB}^H} \mathbf{H}_{\text{eq}}  \mathbf{f}_k^{\text{BB}} s_k + \mathbf{u}_k^{\text{BB}^H} \mathbf{W}^H_{\text{RF}} \mathbf{n} = \rho \mathbf{V}_{k,k}s_k + \mathbf{n'},
    \end{aligned}
\end{equation}
where~$\mathbf{V}_{k,k}=\mathbf{u}_k^{\text{BB}^H} \mathbf{H}_{\text{eq}} \mathbf{f}_k^{\text{BB}}$ represents $k^{th}$ singular value of~$\mathbf{H}_{\text{eq}}$ and also the element at the~$k^{th}$ row and column of the diagonal matrix~$\mathbf{V}$. The vector~$\mathbf{n'} = \mathbf{u}_k^{\text{BB}^H} \mathbf{W}^H_{\text{RF}} \mathbf{n}$ denotes the thermal noise after the analog and digital combining matrices. Therefore, the achievable sum rate can be given as a function of the signal-to-noise ratio~(SNR) at the~$k^{th}$ stream named~$\beta_k$ by
\begin{equation}
 \label{eq:the_se_nopn}
        \text{R} = \sum_{k=1}^{N_{\text{s}}} \text{R}_k = \sum_{k=1}^{N_{\text{s}}} \log_2 \left( 1 + \beta_k \right),
\end{equation}
where~$\beta_k$ is defined by
\begin{equation}
    \begin{aligned}
        \beta_k = \frac{\rho^2 \left\lvert \mathbf{V}_{k,k} \right\rvert^2}{\sigma^2 \xi_k},
    \end{aligned}
    \label{eq:beta_k}
\end{equation}
where~$\xi_k = \mathbf{u}_k^{\text{BB}^H} \mathbf{W}_{\text{RF}}^H \mathbf{W}_{\text{RF}} \mathbf{u}_k^{\text{BB}}$. Using SVD approach given the perfect~CSI knowledge allows to decouple the MIMO channel into~$N_{\text{s}}$ independent SISO channels as highlighted in~\eqref{eq:the_rvd_signal_str2}. In this setting, an exact semi-analytical~BER expression for a~\mbox{$M$-ary QAM} constellation, where~$M$ denotes the modulation order, can be deduced for a given channel assuming an optimal detection as follows~\cite{Kyongkuk,Edward}
\begin{equation}
 \label{eq:ber_nopn}
    \begin{split}
        P_{be} &= \frac{4}{N_{\text{s}} \sqrt{M}  \log_2 (M) } \sum_{k=1}^{N_{\text{s}}} \sum_{p=1}^{\log_2 \sqrt{M} } \sum_{c=0}^{\left(1 -2^{-p}\right)\sqrt{M}-1 } \\
        & \Biggl \lbrace \left(-1\right)^{\left\lfloor \frac{c \cdot 2^{p-1}}{\sqrt{M}}\right\rfloor} \left(2^{p-1} - \left\lfloor \frac{c \cdot 2^{p-1}}{\sqrt{M}} + \frac{1}{2} \right\rfloor \right) \\
        &\mathcal{Q}\left(\sqrt{\frac{3 \left( 2c+1\right)^2 \beta_k}{M-1} } \right) \Biggl \rbrace,
    \end{split}
\end{equation}
where~$\mathcal{Q}\left(\cdot\right)$ represents the Q-function. The semi-analytical BER expression in~\eqref{eq:ber_nopn} depends on the SNR~$\beta_k$ instead of the SNR at the receive antennas (before the analog combining or decoding). Thankfully, the relationship between the two terms is given by
\begin{equation}
\label{eq:sigma_thermal_noise}
    \beta_k = \frac{\left\lvert \mathbf{V}_{k,k} \right\rvert^2 \cdot  \digamma}{\lvert \xi_k \rvert \omega},
\end{equation}
where~$\digamma$ denotes the SNR at the receive antennas. The term~$\omega = \mathbb{E}\left\lbrace \left\lvert \mathbf{h}_q^T \mathbf{F}_{\text{RF}} \mathbf{F}^H_{\text{RF}} \mathbf{h}_q^* \right\rvert \right\rbrace$ with~$\mathbf{H}= \left[\mathbf{h}_1, \cdots, \mathbf{h}_{N_{\text{r}}} \right]^T$.

\subsection{HP Systems with PN}

\subsection*{1. Lower Bound Expression of the Achievable Rate}

The received signal at the $k^{th}$ stream from~\eqref{eq:rcvd_hybp_pn_clo} and assuming the normalized~$\mathbf{F}_{\text{BB}}$ is expressed by
\begin{equation}
        r_k = \rho e^{j\psi} \mathbf{V}_{k,k} s_k + \mathbf{u}_k^{\text{BB}^H} \mathbf{W}^H_{\text{RF}} \mathbf{\tilde{n}},
    \label{eq:rcvd_hybp_pn_clo2_str}
\end{equation}
where~$\mathbf{\tilde{n}} = e^{j\varphi} \mathbf{n} \sim \mathbb{C} \mathcal{N}\left(\mathbf{0},\sigma^2 \mathbf{I}_{N_\text{r}}\right)$ due to the circular symmetric property of $\mathbf{n}$. 

Channel precoding requires channel estimation at the receiver in the downlink, which is then fed back to the transmitter, assuming a sufficient coherence time to perform the precoding. Without PN mitigation, the coherent signal estimated at the received can be expressed by~$\mathbb{E} \left\lbrace e^{j\psi} \mathbf{V}_{k,k} \right \rbrace$. Based on the Bussgang decomposition and the technique used in~\cite{Yu, Pitarokoilis}, the equation~\eqref{eq:rcvd_hybp_pn_clo2_str} can be rewritten as follows
\begin{equation}
    \begin{aligned}
        r_k &= \rho\mathbb{E} \left\lbrace e^{j\psi} \mathbf{V}_{k,k} \right \rbrace s_k + \rho \mathcal{M}_ks_k + \mathbf{u}_k^{\text{BB}^H} \mathbf{W}^H_{\text{RF}} \mathbf{\tilde{n}}, 
    \end{aligned}
    \label{eq:rcvd_hybp_pn_clo2_str2}
\end{equation}
where~$\mathcal{M}_k$ denotes the self-interference defined by
\begin{equation}
      \mathcal{M}_k = e^{j\psi}\mathbf{V}_{k,k} - \mathbb{E} \left\lbrace e^{j\psi} \mathbf{V}_{k,k} \right \rbrace \cdot
    \label{eq:rcvd_hybp_pn_clo2_self_interf}
\end{equation}
The received signal is the sum of uncorrelated terms. The exact probability distribution of~$\mathcal{M}_ks_k$ is difficult to compute. However, its variance can be readily computed under the assumption of perfect CSI. Based on this, we derive a lower bound on the achievable rate by considering the worst-case uncorrelated additive noise as Gaussian with the same variance as~$\rho \mathcal{M}_ks_k + \mathbf{u}_k^{\text{BB}^H} \mathbf{W}^H_{\text{RF}} \mathbf{\tilde{n}}$~\cite{Pitarokoilis}. As a result, the expression of the semi-analytical sum rate as a function of the signal-to-interference-and-noise ratio~(SINR) is obtained as follows
\begin{equation}
        \text{R}_{\text{PN}} = \sum_{k=1}^{N_{\text{s}}} \log_2 \left( 1 + \frac{\rho^2 \left\lvert \mathbb{E} \left\lbrace e^{j\psi}\mathbf{V}_{k,k} \right \rbrace \right\rvert^2 }{ \rho^2 \kappa_k + \mathbb{E} \left\lbrace \left \lvert e^{j\varphi}\mathbf{u}_k^{\text{BB}^H} \mathbf{W}^H_{\text{RF}} \mathbf{n} \right \rvert^2 \right\rbrace} \right),
    \label{eq:the_se_pn_clo}
\end{equation}
where~$\kappa_k = \mathbb{E} \left\lbrace \left \lvert \mathcal{M}_k s_k \right \rvert^2 \right\rbrace$ denotes the interference power. After some computations, the final expression of the achievable rate is given by
\begin{equation}
    \text{R}_{\text{PN}} = \sum_{k=1}^{N_{\text{s}}} \log_2 \left( 1 + \frac{\rho^2 \left\lvert \mathbf{V}_{k,k} \right\rvert^2 }{\rho^2 \left(e^{\sigma^2_\psi} -1 \right) \lvert \mathbf{V}_{k,k} \rvert^2 + \sigma^2 e^{\sigma^2_\psi} \xi_k } \right) \cdot \\
    \label{eq:the_se_pn_clo2}
\end{equation}
\hspace{0.25cm} \textit{Proof}: The proof is given in Appendix~\ref{sec:apdx}-A. $\blacksquare$\\
Then, we deduce the closed-form expression of the lower bound on the achievable~SE in high-SNR regime as
\begin{equation}
    \begin{aligned}   
       \underset{\sigma^2 \rightarrow 0}{\text{R}_{\text{PN}}} &= \sum_{k=1}^{N_{\text{s}}} \log_2 \left( 1 + \frac{\rho^2 \left\lvert \mathbf{V}_{k,k} \right\rvert^2 }{\rho^2 \left(e^{\sigma^2_\psi}-1\right) \lvert \mathbf{V}_{k,k} \rvert^2 } \right) \\
       &= \sum_{k=1}^{N_{\text{s}}} \log_2 \left( 1 + \frac{1}{e^{\sigma^2_\psi}-1} \right) = N_{\text{s}} \log_2 \left(\frac{e^{\sigma^2_\psi} }{e^{\sigma^2_\psi}-1} \right) \cdot  \end{aligned}
    \label{eq:the_se_pn_clo22}
\end{equation}
We observe that the high-SNR approximation of~SE depends only on the~PN power~$\sigma^2_\psi$ and the number of data streams~$N_{\text{s}}$. As a result, higher~PN degrades~SE while increasing~$N_{\text{s}}$ improves the latter for a given~GPN regime.

\subsection*{2. Performance Improvement with PN Mitigation}

Since we assume a perfect CSI knowledge, one can transmit pilot on few streams every time to track and compensate the PN effect at the receiver. Hence, the received signal in~\eqref{eq:rcvd_hybp_pn_clo2_str} after PN mitigation is expressed as
\begin{equation}
    \begin{aligned}
        \tilde{r}_k &= \rho \mathbf{V}_{k,k} s_k +  e^{-j\hat{\psi}}\mathbf{u}_k^{\text{BB}^H} \mathbf{W}^H_{\text{RF}} \mathbf{\tilde{n}} \\ 
        &= \rho \mathbf{V}_{k,k} s_k +  \mathbf{u}_k^{\text{BB}^H} \mathbf{W}^H_{\text{RF}} \mathbf{n'},
    \end{aligned}
    \label{eq:rcvd_hybp_nopn_clo_str}
\end{equation}
where~$\mathbf{n'} = e^{-j\hat{\psi}}\mathbf{\tilde{n}} \sim \mathbb{C} \mathcal{N}\left(\mathbf{0},\sigma^2 \mathbf{I}_{N_\text{r}}\right)$. The term~$\hat{\psi}$ represents the estimated PN which is given by
\begin{equation}
    \begin{aligned}
        \hat{\psi} &= \text{arg} \left\lbrace \frac{1}{N_{\text{pil}}} \sum_{q} \frac{r_q s^*_q}{\rho \mathbf{V}_{q,q} \left\lvert s_q\right\rvert^2}\right\rbrace,
    \end{aligned}
    \label{eq:rcvd_hybp_nopn_clo_pn_estim}
\end{equation}
where~$s_q$ and~$N_{\text{pil}}$ are the transmitted signal at the~$q^{th}$ stream and the number of streams allocated for PN tracking, respectively. One can notice that after PN compensation,~\eqref{eq:rcvd_hybp_nopn_clo_str} and~\eqref{eq:the_rvd_signal_str2} are similar. According to~\eqref{eq:the_se_nopn}, the semi-analytical sum rate is defined by
\begin{equation}
 \label{eq:the_se_clo_pn_compensated}
    \begin{aligned}
        \text{R} =\sum_{k=1}^{N} \log_2 \left( 1 + \beta_k \right),
    \end{aligned}
\end{equation}
where~$N = N_{\text{s}} - N_{\text{pil}}$. Besides, the~BER of an SVD-based HP SU-MIMO system employing~$M$-ary square QAM after PN mitigation can be approximated to the one given in~\eqref{eq:ber_nopn} assuming a good PN estimation. For a SVD-based MIMO system, a very small~$N_{\text{pil}}$ may suffice to counteract PN thanks to its SNR maximization benefit. Consequently, inserting pilots might be an optimal approach for achieving high data rate when employing high~$N_{\text{s}}$ value (w.r.t. the condition $N_{\text{s}} \leq N_{\text{RF}} \ll N_{\text{t}}$). However, since PN tracking requires pilot insertion, this will reduce the~SE~(especially for low~$N_{\text{s}}$ values). 

\subsection*{3. Performance Improvement without PN Compensation: Optimal Receiver and Modulation Scheme}

\begin{figure*}[t]
   \centering   
   \begin{subfigure}[b]{0.45 \textwidth}
       \centering
       \includegraphics[width=0.9\columnwidth]{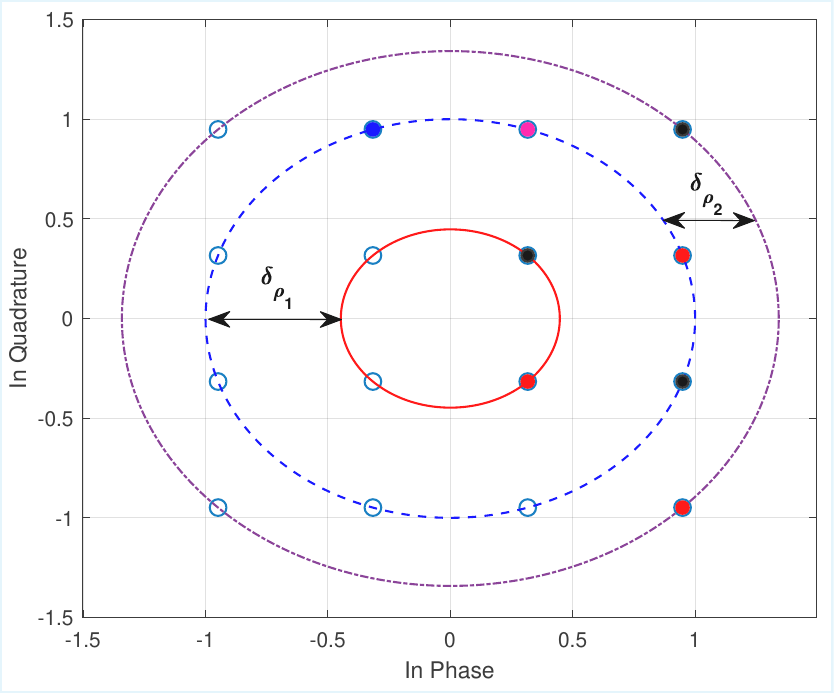}
       \caption{Cartesian domain}
       \label{fig:16_QAM_cartesian}
   \end{subfigure}
    \begin{subfigure}[b]{0.45 \textwidth}
       \centering
       \includegraphics[width=0.9\columnwidth]{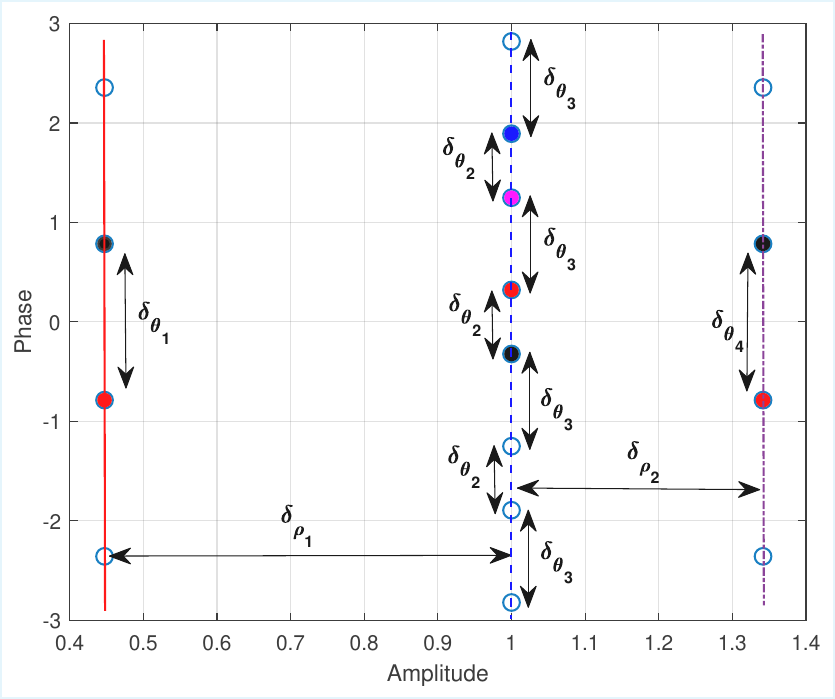}
       \caption{Polar domain}
       \label{fig:16_QAM_polar}
   \end{subfigure}
   \caption{Representation of the 16-QAM constellation in cartesian and polar domains.}
    \label{fig:16_QAM_representation}
    \vspace{-0.3cm}
\end{figure*}

Assuming perfect CSI knowledge, the expression~\eqref{eq:rcvd_hybp_pn_clo2_str} after normalizing by the factor~$\rho \mathbf{V}_{k,k}$ becomes 
\begin{equation}
    \begin{aligned}
        \tilde{r}_k &= s_k e^{j\psi} +  \frac{e^{j\varphi}}{\rho \mathbf{V}_{k,k}}\mathbf{u}_k^{\text{BB}^H} \mathbf{W}^H_{\text{RF}} \mathbf{\tilde{n}} \\
        &= s_k e^{j\psi} + n'_k,
    \end{aligned}
    \label{eq:rcvd_hybp_pn_clo_str_normalize}
\end{equation}
where~$n'_k \sim \mathbb{C} \mathcal{N}\left(\mathbf{0},2\sigma_{n_k}^2 \right)$ with~$\sigma_{n_k}^2$ given by
\begin{equation}
    \sigma_{n_k}^2 = \frac{1}{2\beta_k},
    \label{eq:sigma2_n}
\end{equation}
where~$\beta_k$ is defined in~\eqref{eq:beta_k}. One can notice through~\eqref{eq:rcvd_hybp_pn_clo_str_normalize} how the~SVD-based HP SU-MIMO systems using single-carrier~(SC) modulation at the~$k^{th}$ stream is equivalent to a~SC single-input-single-output system model in the presence of GPN $\psi$ and additive complex Gaussian noise~$\tilde{n}_k$. Optimal detectors such as polar metric (PM) have been proposed in the literature to improve data detection and thus achieve high throughput under strong GPN conditions~\cite{Bicaïs2}.

For symbol-by-symbol detection with equiprobable and independent symbols, the maximum likelihood criterion minimizes the symbol error probability. The likelihood function can thus be expressed as follows
\begin{equation}
    p(r_k|s_k) = p(r_{k_\rho}, r_{k_\theta} | s_{k_\rho}, s_{k_\theta}) \cdot
\end{equation}
The expression~\eqref{eq:rcvd_hybp_pn_clo_str_normalize} can be rewritten as follows
\begin{equation}
    \begin{aligned}
        \tilde{r}_k = s_k e^{j\psi} + n'_k &= s_{k_\rho}e^{j\left(s_{k_\theta} + \psi \right)} + n'_k \\
        &= \left(s_{k_\rho} + n''_k \right)e^{j\left(s_{k_\theta} + \psi \right)},
    \end{aligned}
    \label{eq:rcvd_hybp_pn_clo_str_normalize2}
\end{equation}
where~$n''_k = n'_k e^{-j\left(s_{k_\theta} + \psi \right)} \sim \mathbb{C} \mathcal{N}\left(\mathbf{0},2\sigma_{n_k}^2 \right)$. Sub-THz targets high rates and massive MIMO systems allows high-SNR assuming high antenna gains. Consequently, the amplitude of the received symbol at the $k^{th}$ stream can be given assuming high SNR approximation by~\cite{Bicaïs}
\begin{equation}
        \tilde{r}_{k_\rho} = \left \lvert \left(s_{k_\rho} + n''_k \right)e^{j\left(s_{k_\theta} + \psi \right)} \right \rvert \simeq s_{k_\rho} + \mathfrak{Re}\left\lbrace n''_k \right \rbrace,
    \label{eq:rcvd_hybp_pn_clo_str_magnitude}
\end{equation}
and the phase by
\begin{equation}
        \tilde{r}_{k_\theta} = \text{arg} \left \lbrace \left(s_{k_\rho} + n''_k \right)e^{j\left(s_{k_\theta} + \psi \right)} \right \rbrace \simeq s_{k_\theta} + \psi + \frac{\mathfrak{Im}\left\lbrace n''_k \right \rbrace}{s_{k_\rho}} \cdot
    \label{eq:rcvd_hybp_pn_clo_str_phase}
\end{equation}
Under this condition, it directly follows from the channel and~PN models that
\begin{equation}
    \begin{aligned}
    &\tilde{r}_{k_\rho} - s_{k_\rho} \sim \mathcal{N} \left( 0, \sigma^2_n\right) \\
    &\tilde{r}_{k_\theta} - s_{k_\theta} \sim \mathcal{N} \left( 0, \sigma^2_\psi + \sigma_{n_k}^2 / s^2_{k_\rho} \right) \cdot
    \end{aligned}
    \label{eq:rcvd_hybp_pn_clo_str_distr}
\end{equation}
Therefore, the channel likelihood function can be expressed as a bivariate Gaussian distribution
\begin{equation}
    p(\tilde{r}_k|s_k) = \frac{\text{exp} \left( -\frac{1}{2}\left(\frac{\left(\tilde{r}_{k_\rho} - s_{k_\rho}\right)^2}{\sigma_{n_k}^2} + \frac{\left(\tilde{r}_{k_\theta} - s_{k_\theta}\right)^2}{\sigma^2_\psi + \sigma_{n_k}^2 / s^2_{k_\rho}}\right)\right)}{2\pi \sqrt{\sigma^2_{n_k}\left(\sigma^2_\psi + \sigma_{n_k}^2 / s^2_{k_\rho}\right)}} \cdot
    \label{eq:pdf_str}
\end{equation}
From~\eqref{eq:pdf_str}, we deduce the PM whose decision rule is given by~\cite{Bicaïs}
\begin{equation}
    \hat{s}_k = \underset{s \in \mathcal{C}} {\text{argmin}} \; d_\gamma \left(\tilde{r}_k,s\right),
    \label{eq:rcvd_hybp_pn_clo_str_decis_rule}
\end{equation}
where~$\mathcal{C}$ is the set containing the reference symbols of chosen modulation scheme. The term~$d_\gamma$ standing for the PM is defined by
\begin{equation}
    d_\gamma \left(\tilde{r}_k,s\right) =  \frac{\left(\tilde{r}_{k_\rho} - s_\rho\right)^2}{\sigma_{n_k}^2} + \frac{\left(\tilde{r}_{k_\theta} - s_\theta\right)^2}{\gamma^2},
    \label{eq:rcvd_hybp_pn_clo_str_pm}
\end{equation}
with~$\gamma^2 = \sigma^2_\psi + \sigma_{n_k}^2/E_s = \sigma^2_\psi + \sigma_{n_k}^2$ assuming normalized modulation scheme~(e.g., QAM symbols), \textit{i.e.},~$E_s = \mathbb{E} \left \lbrace \left \lvert s \right \rvert^2\right \rbrace=1$. By considering equiprobable and independent symbols, we can define the detection error probability~$P_{e_k}$ at the~$k^{th}$ stream by
\begin{equation}
    P_{e_k} = \frac{1}{M} \sum_{s \in \mathcal{C}} P\left(\hat{s}_k \neq s_k | s\right) = \frac{1}{M} \sum_{s \in \mathcal{C}} \left(1-P\left(\hat{s}_k = s_k | s\right)\right),
    \label{eq:Pe_total}
\end{equation}
where the probability of detecting properly according to the polar domain can be approximated by
\begin{equation}
    \begin{split}
      P\left(\hat{s}_k = s_k | s\right) \simeq & P\left(-\frac{\delta_\rho}{2}< \tilde{r}_{k_\rho} - s_\rho < \frac{\delta_\rho}{2} \right) \\
      & \times P \left(-\frac{\delta_\theta}{2}< \tilde{r}_{k_\theta} - s_\theta < \frac{\delta_\theta}{2} \right) \Biggl)\cdot
    \end{split}
\end{equation}
Fig.~\ref{fig:16_QAM_polar} shows the polar representation of the 16-QAM reference symbols. One can notice three magnitude levels: red, blue and purple. For the 16-QAM modulation, we have

\begin{equation}
      \begin{aligned}
      \left\{
    \begin{array}{ll}
        \delta_{\rho_1} = 1 - \left(1/\sqrt{5}\right)\;, \; \delta_{\rho_2} = \left(3/\sqrt{5}\right)-1  , \; \delta_{\theta_1}=\delta_{\theta_4} = \pi/2 \\
       \delta_{\theta_2} = 2 \cdot \arctan\left(\frac{1}{\sqrt{M}-1}\right)= 2 \cdot \arctan\left(\frac{1}{3}\right)\\
       \delta_{\theta_3} = \arctan \left(\sqrt{M}-1\right) - \delta_{\theta_2}/2 = \arctan(3) - \delta_{\theta_2}/2, 
    \end{array}
\right.
     \end{aligned} 
     \label{eq:delta_rho_theta}
\end{equation}
where~$\delta_{\rho_1}>\delta_{\rho_2}$ and $\delta_{\theta_3}>\delta_{\theta_2}$. Regarding the angle symmetry of QAM constellation points, we can approximate the total detection error probability of the~$16$-QAM from~\eqref{eq:Pe_total} by
\begin{equation}
    \begin{aligned}   
      P_{e_k} & \simeq \frac{1}{2} \left(\mathcal{Q}\left(\frac{\delta_{\rho_1}}{2\sigma_{n_k}}\right) + 3 \mathcal{Q}\left(\frac{\delta_{\rho_2}}{2\sigma_{n_k}}\right)\right) \\ 
      &+ \mathcal{Q}\left(\frac{\delta_{\theta_1}}{2\sqrt{ \sigma^2_\psi  + \sigma_{n_k}^2}}\right) + \mathcal{Q}\left(\frac{\delta_{\theta_2}}{2\sqrt{ \sigma^2_\psi  + \sigma_{n_k}^2}}\right) \cdot
    \end{aligned}
    \label{eq:Pe_16_QAM}
\end{equation}
\hspace{0.25cm} \textit{Proof}: The proof is given in Appendix~\ref{sec:apdx}-B. $\blacksquare$\\
By replacing~\eqref{eq:sigma2_n} in~\eqref{eq:Pe_16_QAM}, we obtain the~BER expression at the~$k^{th}$ stream defined by
\begin{equation}
    \begin{aligned}       
      P_{e_k} & \simeq \frac{1}{2} \left( \mathcal{Q}\left(\sqrt{\frac{\delta^2_{\rho_1} \beta_k}{2}}\right) + 3 \mathcal{Q}\left(\sqrt{\frac{\delta^2_{\rho_2} \beta_k }{2}}\right) \right) \\
      &+ \mathcal{Q}\left(\frac{\delta_{\theta_1}}{2\sqrt{ \sigma^2_\psi  + \frac{1}{2 \beta_k}}}\right) + \mathcal{Q}\left(\frac{\delta_{\theta_2}}{2\sqrt{ \sigma^2_\psi  + \frac{1}{2 \beta_k}}} \right),
    \end{aligned}
    \label{eq:Pe_16_QAM2}
\end{equation}
where~$\delta_{\rho_1}, \delta_{\rho_2}, \delta_{\theta_1}$~and~$\delta_{\theta_2}$ are given in~\eqref{eq:delta_rho_theta}. The expression of the error probability as a function of the SNR at the~Rx antennas can be deduced by replacing~\eqref{eq:sigma_thermal_noise} in~\eqref{eq:Pe_16_QAM2}. Finally, for a given channel, the semi-analytical BER expression of SVD-based HP SU-MIMO systems employing the~16-QAM and assuming the CLO scheme is expressed by
\begin{equation}    
     \begin{aligned}
        P_{be} & \simeq \frac{1}{N_{\text{s}}\log_2(M)} \sum_{k=1}^{N_{\text{s}}} P_{e_k} \simeq \frac{1}{4N_{\text{s}}} \sum_{k=1}^{N_{\text{s}}} P_{e_k},
      \end{aligned}
    \label{eq:BER_16_QAM}
\end{equation}
with~$P_{e_k}$ defined in~\eqref{eq:Pe_16_QAM2}. The first two terms represent the probability of incorrectly estimating the amplitude level, while the last two terms capture the probability of erroneously detecting the phase of the received signal. When the PN becomes strong, the BER reaches an error floor which is invariant despite the increasing SNR. According to the~\mbox{$16$-QAM}, this PN-induced error floor, when the~SNR tends to infinity~(\textit{i.e.},~$\beta_k \rightarrow \infty$), can be evaluated by
\begin{equation}
    \begin{aligned}       
      \underset{\beta_k \rightarrow \infty}{P_{be}} \simeq \frac{1}{4} \left(\mathcal{Q}\left(\frac{\delta_{\theta_1}}{2\sqrt{ \sigma^2_\psi}}\right) + \mathcal{Q}\left(\frac{\delta_{\theta_2}}{2\sqrt{ \sigma^2_\psi}}\right) \right) \cdot
    \end{aligned}
    \label{eq:Pe_16_QAM_final}
\end{equation}
One can notice in~\eqref{eq:Pe_16_QAM_final} that the average BER only depends on the PN variance in high SNR regime. This allows to consider~\eqref{eq:Pe_16_QAM_final} as a closed-form expression of the average BER using the~\mbox{16-QAM}. It is important to mention that~\eqref{eq:Pe_16_QAM_final} is more accurate in strong GPN scenario and high SNR regime as highlighted in the numerical results section. In the medium PN regime, this BER expression can serve as a lower bound when the Euclidean detector is employed.
 \begin{figure}[tb]
   \centering    
    \includegraphics[width=1\columnwidth]{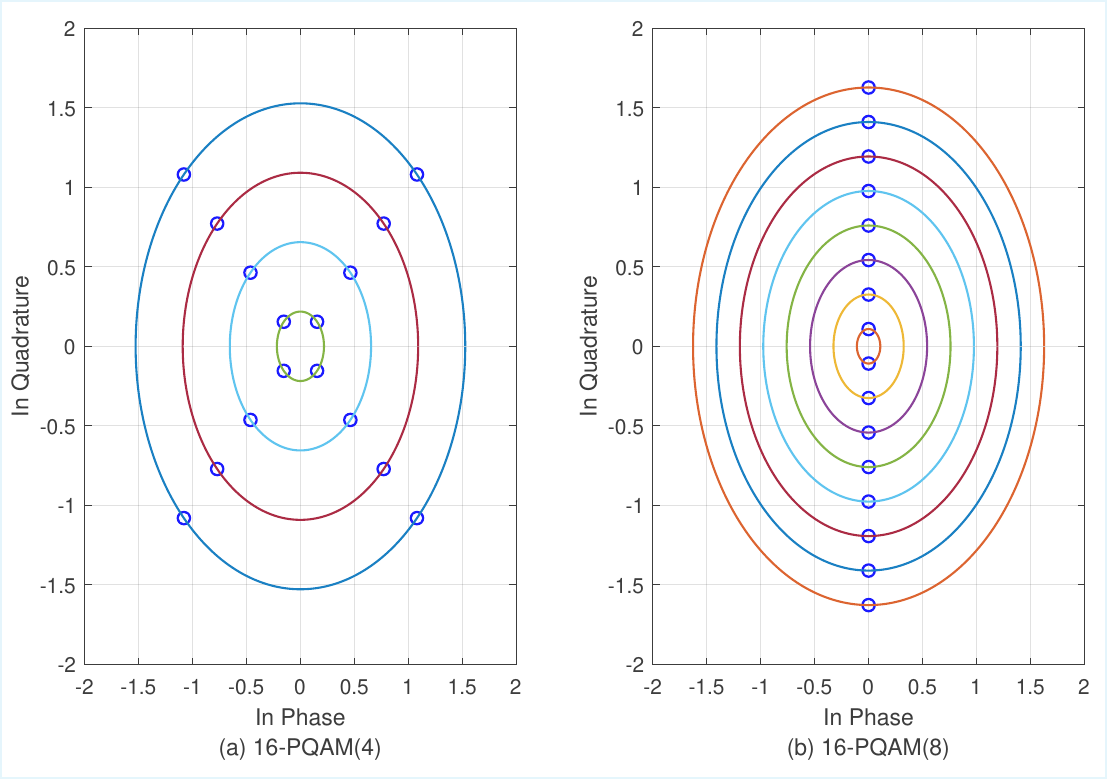}
    \caption{Reference constellation scheme of a 16-PQAM($\Gamma$) modulation considering $\Gamma = \lbrace 4,8\rbrace$.} 
    \label{fig:contel_16PQAM}
    \vspace{-0.2cm}
\end{figure}

\begin{figure}[tb]
   \centering    
    \includegraphics[width=0.85\columnwidth]{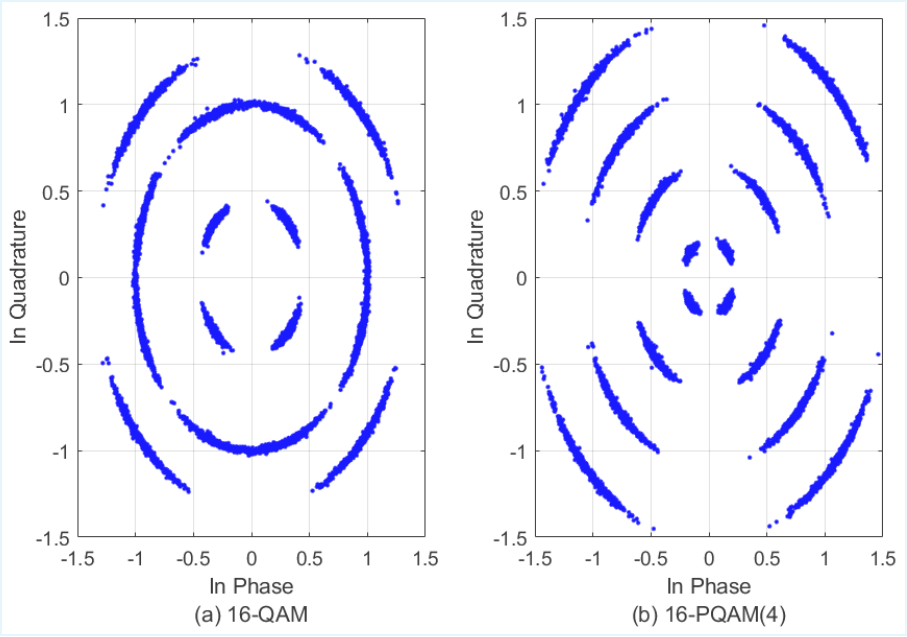}
    \caption{Received signal constellation for both 16-QAM and 16-PQAM(4) in the medium GPN. We consider a SNR of $30$ dB, $N_{\text{RF}}=N_{\text{s}}=4$, $N_{\text{t}}=144$, $N_{\text{r}}=36$.}
    \label{fig:contel_16QAM_16PQAM__clo}
    \vspace{-0.4cm}
\end{figure}

Polar modulations such as PQAM have been proposed to maximize the signal detection accuracy~\cite{Bicaïs}. This new modulation, created for~SC systems strongly impaired by the PN, maximizes the phase distance between the constellation points in order to avoid detection errors.~\mbox{$M$-PQAM($\Gamma$)} denotes the use of the PQAM scheme where $M$ is the modulation order. The $\Gamma$ value stands for the shape which represents the number of amplitude level. Further, a \mbox{$M$-PQAM($M/2$)} describes an amplitude-shift keying while \mbox{$M$-PQAM(1)} is a phase-shift keying. Fig.~\ref{fig:contel_16PQAM} illustrates the PQAM for $M=16$ and $\Gamma=\lbrace 4,8 \rbrace$, and Fig.~\ref{fig:contel_16QAM_16PQAM__clo} shows the received signal constellation when~\mbox{16-QAM} and~\mbox{16-PQAM(4)} are used under medium GPN regime. Intuitively to the SC model~\cite{Bicaïs}, the semi-analytical BER for a given channel matrix, when the~\mbox{$M$-PQAM($\Gamma$)} constellation is considered, can be approximated by the following closed-form expression
\begin{equation}
 \label{eq:ber_clo_pn_pqam}
    \begin{split}
        P_{e_k} & \simeq \frac{2}{\log_2(M)} \Biggl( \mathcal{Q}\left(\sqrt{\frac{6 \left \lvert \mathbf{V}_{k,k} \right\rvert^2 \digamma }{\left(4\Gamma^2-1\right)\omega \left \lvert \xi_k\right\rvert} } \right)  \\
        &+ \mathcal{Q}\left( \frac{\pi \Gamma }{M\sqrt{\sigma^2_\psi + \frac{\omega \left \lvert \xi_k\right\rvert}{2\left \lvert \mathbf{V}_{k,k} \right\rvert^2  \digamma} }} \right) \Biggr) \cdot
    \end{split}
\end{equation}
\hspace{0.25cm} \textit{Proof}: The proof is given in Appendix~\ref{sec:apdx}-C. $\blacksquare$\\
The first and second term express the probability of misestimating the amplitude level and the probability of misestimating the phase of the received signal, respectively. Similarly to the QAM case, the closed-form expression of the average BER in high-SNR and strong~GPN regimes can be approximated by
\begin{equation}
    \begin{aligned}       
      \underset{\digamma \longrightarrow \infty}{P_{be}} \simeq \frac{2}{\log_2(M)} \mathcal{Q}\left( \frac{\pi \Gamma }{M\sqrt{\sigma^2_\psi }} \right) \cdot
    \end{aligned}
    \label{eq:ber_clo_pn_pqam2}
\end{equation}
By substituting~$\Gamma=1$ and~$M=4$ into~\eqref{eq:ber_clo_pn_pqam}, we can deduce the semi-analytical BER expression of~$4$-QAM. Moreover, the PN-induced error floor expression for~\mbox{4-QAM}, assuming~high-SNR and strong GPN regimes, is obtained by substituting \mbox{$\Gamma=1$} and~$M=4$ into~\eqref{eq:ber_clo_pn_pqam2} as follows
\begin{equation}     
      \underset{\digamma \longrightarrow \infty}{P_{be}} \simeq \mathcal{Q}\left( \frac{\pi }{4\sqrt{\sigma^2_\psi }} \right) \cdot
    \label{eq:ber_clo_pn_4qam}
\end{equation}
Increasing the value of~$\Gamma$ provides robustness to the PN but also an additive noise weakness. Thereby, the PQAM necessitates a high SNR as depicted in Fig.~\ref{fig:contel_16QAM_16PQAM__clo}b for enhancing the system performance when GPN becomes important~($\sigma^2_\psi \geq 10^{-2}$ for $M \geq 16$). For low PN regime (depending on the modulation order), the conventional QAM is more suitable for transmission. The derivation process of the analytical BER for $M$-ary QAM constellation can be employed for~$M \geq 16$. In this work, we only highlight the closed-form BER expression of~\mbox{4-QAM} and~\mbox{16-QAM} because the higher the modulation order, the greater the~PN impact on system performance.

It is important to mention that the pure rotation of the transmitted symbols illustrated in~Fig.~\ref{fig:contel_16QAM_16PQAM__clo} is only possible when a SC waveform is considered. In the case of a multi-carrier~(MC) waveform such as OFDM, the presence of PN induces the common phase error and intercarrier interference~\cite{Afshang}. Thus, this will break the pure rotation observed and thus, making the PQAM with the PM useless. Fortunately, the detection metric proposed in~\cite{Bello2} could be a solution to enhance the performance under GPN influence if DFT-s-OFDM waveform is used.  

\section{Numerical Results} \label{sec:results}

\subsection{Simulation Environment}
In this section, we highlight the simulation results of HP MIMO systems with GPN impairments. For instance, we show the GPN effect on both achievable SE and uncoded BER performance as a function of the SNR at the~Rx antennas. For the BER performance, we choose a target BER of $10^{-4}$. As a remainder, three GPN levels are considered: strong GPN for $\sigma^2_\psi = 10^{-1}$, medium GPN for $\sigma^2_\psi = 10^{-2}$ and low GPN for $\sigma^2_\psi = 10^{-3}$ assuming $\sigma^2_\phi = \sigma^2_\varphi$. The channel parameters are set as $N_\text{c}=5$ clusters and $N_\text{R}=10$ rays per cluster. Similar to~\cite{Lin}, we assume the AoA and AoD follow the Laplacian distribution with uniformly distributed mean angles over $[0, 2\pi)$ and angular spread of $10$ degrees. The simulation results are averaged over $2 \times 10^5$ channel realizations. For the analog precoder~$\mathbf{F}_{\text{RF}}$ and decoder design~$\mathbf{W}_{\text{RF}}$, we consider the PE-AltMin algorithm proposed in~\cite{Yu}. For the digital precoder~$\mathbf{F}_{\text{BB}}$ and decoder~$\mathbf{W}_{\text{BB}}$, we compute them from the SVD of the equivalent channel~$\mathbf{H}_{\text{eq}}=\mathbf{W}^H_{\text{RF}}\mathbf{H}\mathbf{F}_{\text{RF}}$. FDP stands for full digital precoding and when nothing is mentioned,~the HP is considered. The terms~\mbox{PM-D} and~\mbox{EUC-D} respectively stand for the implementation of the~PM detector~\footnote{Obviously, we assume the perfect knowledge of GPN variance $\sigma_\psi^2$ and noise variance~$\sigma_{n_k}^2$ which can be estimated following~\cite{Bicaïs} and assuming CSI knowledge.} and~Euclidean detector~(EUC-D) whose decision rule is defined by
\begin{equation}
    \hat{s}_k = \underset{s \in \mathcal{C}} {\text{argmin}} \; \parallel \tilde{r}_k - s\parallel^2 \cdot
    \label{eq:rcvd_hybp_pn_ilo_str_decis_rule}
\end{equation}
 \begin{figure}[tb]
   \centering    
    \includegraphics[width=0.85\columnwidth]{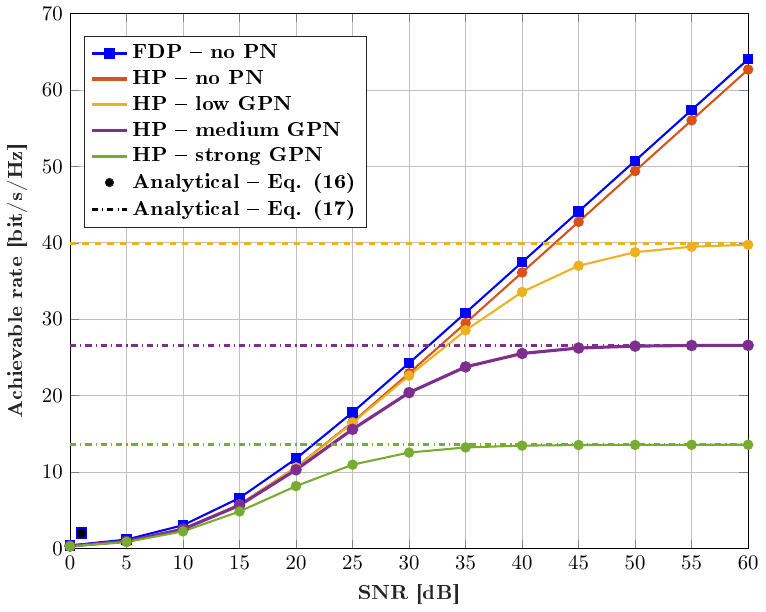}
    \caption{SE performance as a function of the SNR considering different GPN levels. We use~$N_{\text{RF}}=N_{\text{s}}=4$, $N_{\text{t}}=144$, $N_{\text{r}}=36$.}
    \label{fig:SE_ilo_vs_clo}
   \vspace{-0.3cm}
\end{figure}

 \begin{figure}[tb]
   \centering    
    \includegraphics[width=0.85\columnwidth]{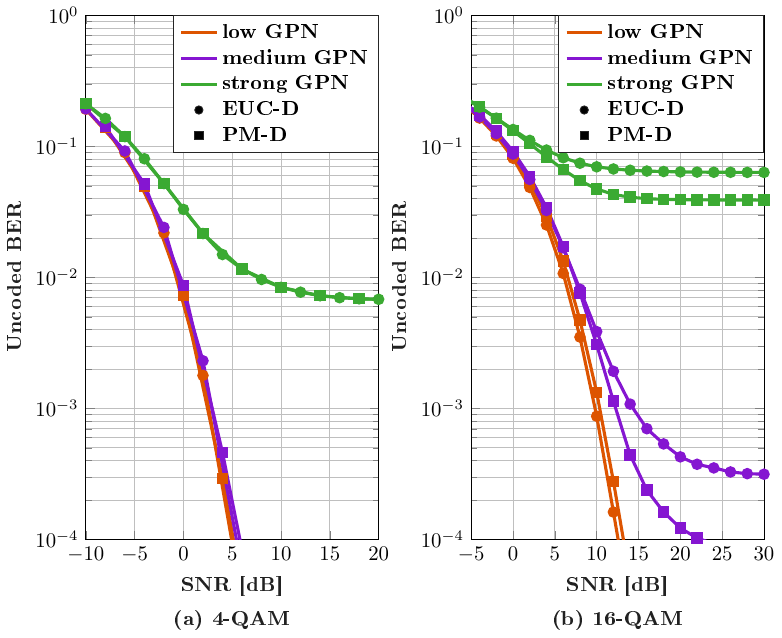}
    \caption{Monte Carlo BER performance comparison between EUC-D and PM-D as a function of the SNR for different GPN regimes. We consider~$N_{\text{RF}}=N_{\text{s}}=4$, $N_{\text{t}}=144$, $N_{\text{r}}=36$.}
    \label{fig:BER_euc_vs_pm_4QAM_16QAM}
   \vspace{-0.3cm}
\end{figure}

 \begin{figure}[tb]
   \centering    
    \includegraphics[width=0.865\columnwidth]{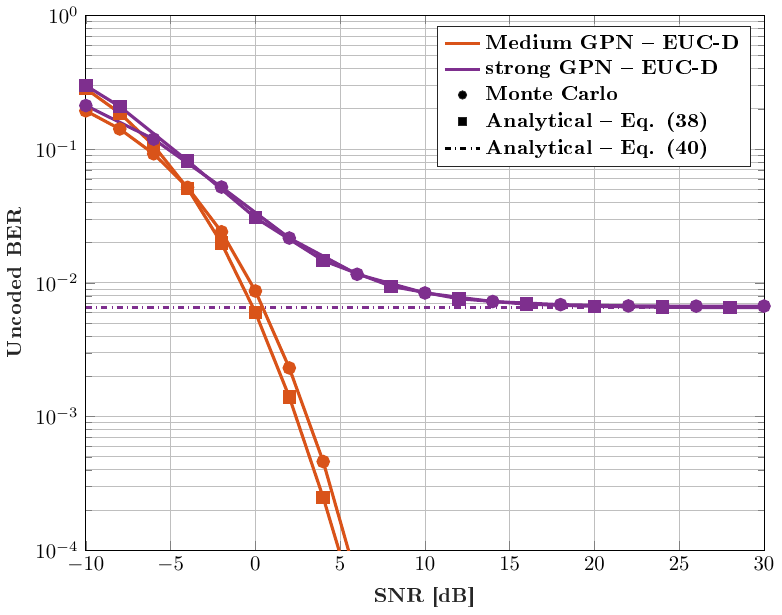}
    \caption{Monte Carlo vs. Analytical in terms of uncoded BER considering 4-QAM. We consider~$N_{\text{RF}}=N_{\text{s}}=4$, $N_{\text{t}}=144$, $N_{\text{r}}=36$.}
    \label{fig:BER_monte_carlo_vs_theoretical_4QAM}
    \vspace{-0.3cm}
\end{figure}

 \begin{figure}[tb]
   \centering    
    \includegraphics[width=0.86\columnwidth]{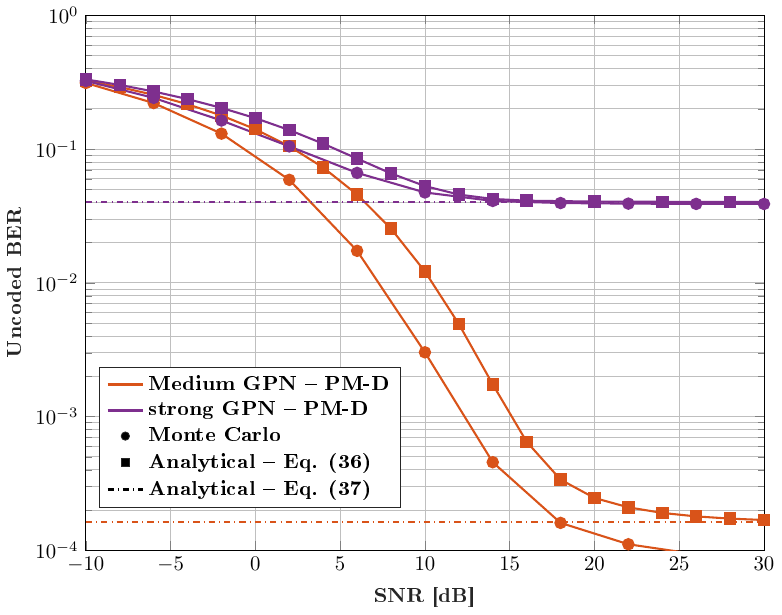}
    \caption{Monte Carlo vs. Analytical in terms of uncoded BER considering 16-QAM. We consider~$N_{\text{RF}}=N_{\text{s}}=4$, $N_{\text{t}}=144$, $N_{\text{r}}=36$.}
    \label{fig:BER_monte_carlo_vs_theoretical_16QAM}
    \vspace{-0.3cm}
\end{figure}

\subsection{Results}
\subsection*{1. Achievable SE Performance} 

Fig.~\ref{fig:SE_ilo_vs_clo} shows the achievable SE performance comparison for different GPN levels. Firstly, we can notice the performance of HP considering the PE-AltMin almost matches the one of FDP when the~PN is missing. This result is similar to the one presented in~\cite{Yu}. Indeed, when~PN is absent, the lower bound expression in~\eqref{eq:the_se_pn_clo2} becomes a semi-analytical sum rate similarly to~\eqref{eq:the_se_nopn}. Secondly, we observe that the more the GPN variance, the more the SE deterioration. Additionally, we remark that the performance using the closed form expression~\eqref{eq:the_se_pn_clo22} matches well the one using the semi-analytical expression~\eqref{eq:the_se_pn_clo2} in high-SNR regime. Moreover, the stronger the PN the lower the SNR validating the closed-form expression of the lower bound on the achievable SE in high-SNR regime.

\subsection*{2. Uncoded BER Performance} 

Fig.~\ref{fig:BER_euc_vs_pm_4QAM_16QAM}a and Fig.~\ref{fig:BER_euc_vs_pm_4QAM_16QAM}b highlight the uncoded BER performance comparison between the EUC-D and the PM-D for different GPN regimes considering a \mbox{4-QAM} and \mbox{16-QAM} modulation, respectively. In Fig.~\ref{fig:BER_euc_vs_pm_4QAM_16QAM}a, we remark that the system reaches the target~BER in low and medium GPN regimes. Nevertheless, the system does not achieve the target BER in strong GPN regime and present an error floor. Further, the~\mbox{PM-D} provides similar performance as the EUC-D and thus, making the~\mbox{PM-D} not necessary for low modulation order. In Fig.~\ref{fig:BER_euc_vs_pm_4QAM_16QAM}b, we notice that the system performance when~\mbox{16-QAM} is used, does not reach the target BER in the medium GPN. Globally, the PN impacts more the system performance with~\mbox{16-QAM} unlike~\mbox{4-QAM}. 

In fact, increasing the modulation order amplifies the noise sensitivity and therefore more detection errors. Moreover, one can remark that the performance considering the~\mbox{PM-D} outperforms the one  using the~\mbox{EUC-D} in both strong and medium~GPN whereas the opposite is noticeable for the low~GPN regime. However, the system performance does not achieve the target~BER even with the~\mbox{PM-D} in the strong GPN level. As a result, the GPN effect depends on the modulation order employed. The higher the modulation order, the stronger the GPN impairments.
 \begin{figure}[tb]
   \centering    
    \includegraphics[width=0.85\columnwidth]{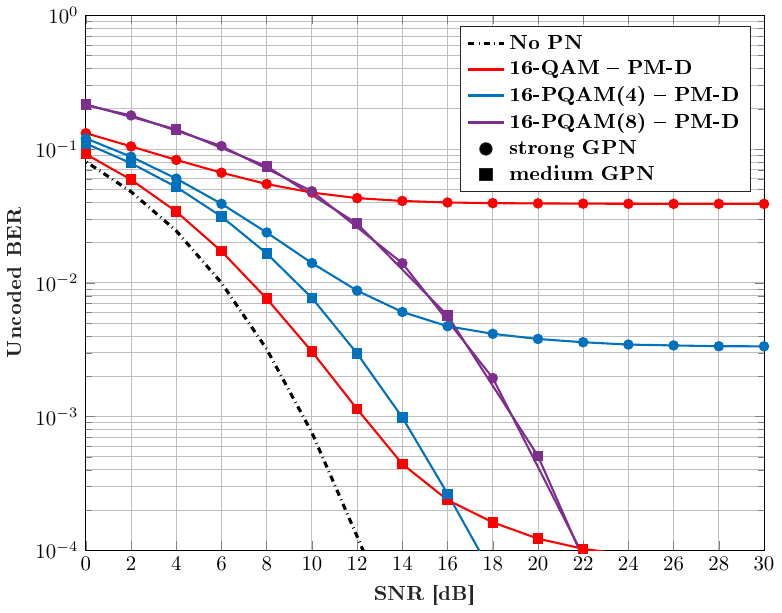}
    \caption{BER performance comparison between QAM and PQAM constellations as a function of the SNR under medium and strong GPN regimes. We use~$N_{\text{RF}}=N_{\text{s}}=4$, $N_{\text{t}}=144$, $N_{\text{r}}=36$.}
    \label{fig:BER_qam_vs_pqam_pm}
    \vspace{-0.3cm}
\end{figure}

 \begin{figure}[tb]
   \centering    
    \includegraphics[width=0.8\columnwidth]{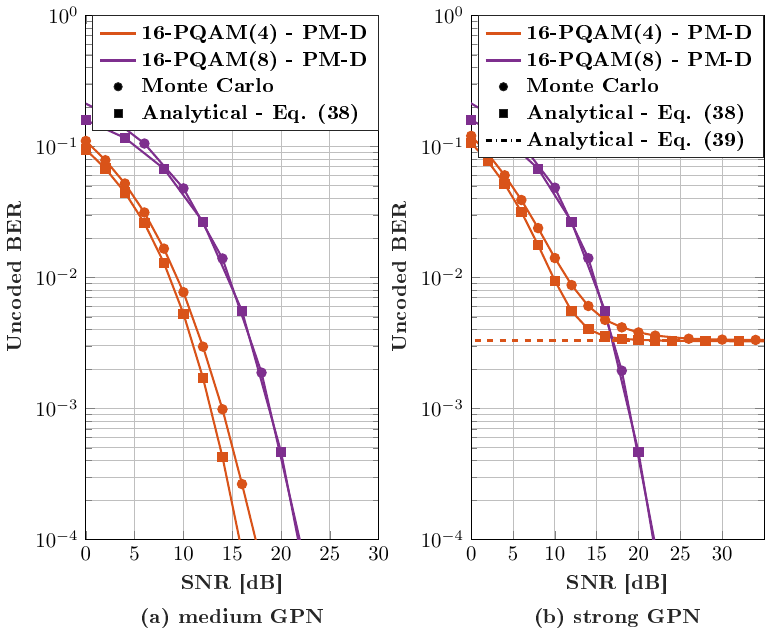}
    \caption{Monte Carlo vs. Analytical in terms of uncoded BER using both $16$-PQAM($4$) and $16$-PQAM($8$) under medium and strong GPN regimes. We consider~$N_{\text{RF}}=N_{\text{s}}=4$, $N_{\text{t}}=144$, $N_{\text{r}}=36$.}
    \label{fig:BER_monte_carlo_vs_theoretical_PQAM}
   \vspace{-0.3cm}
\end{figure}

Fig.~\ref{fig:BER_monte_carlo_vs_theoretical_4QAM} and Fig.~\ref{fig:BER_monte_carlo_vs_theoretical_16QAM} compare respectively Monte Carlo simulation results with the analytical expressions from the previous section. For \mbox{4-QAM}, the Monte Carlo performance perfectly matches the analytical expression (with $M=4$ and $\Gamma=1$ substituted in~\eqref{eq:ber_clo_pn_pqam}) in strong GPN regime. Furthermore, we observe that the performance using the closed-form expression~\eqref{eq:ber_clo_pn_4qam} matches well the one using the semi-analytical expression~\eqref{eq:ber_clo_pn_pqam} in the high-SNR regime under strong GPN. This substantiates the closed-form BER expression derived. In the medium GPN regime, one can notice a slightly difference between the Monte carlo with the semi-analytical BER. 

Regarding the~\mbox{16-QAM}, the Monte Carlo performance matches with the semi-analytical one under strong GPN level. Moreover, the closed-form BER expression in~\eqref{eq:Pe_16_QAM_final} perfectly coincides the Monte Carlo in high-SNR regime and thus validates the latter. Despite the good matching between~\eqref{eq:BER_16_QAM} and~\eqref{eq:Pe_16_QAM_final} in high-SNR, Monte Carlo performance outperforms the one with the semi-analytical BER expression~\eqref{eq:BER_16_QAM} in medium GPN level. Nevertheless, the close-form BER expression given in~\eqref{eq:Pe_16_QAM_final} can be seen as an upper bound BER performance in medium GPN.

Fig.~\ref{fig:BER_qam_vs_pqam_pm} shows the Monte Carlo performance comparison between~QAM and~PQAM constellation schemes. We perform the comparison by considering: (i)~\mbox{16-QAM} modulation associated to the~\mbox{PM-D} only~\footnote{We only consider the~\mbox{PM-D} because we highlight the outperformance of the~\mbox{PM-D} compared to the~\mbox{EUC-D} in medium and strong GPN regimes in Fig.~\ref{fig:BER_euc_vs_pm_4QAM_16QAM}.} for the signal demodulation, and (ii) \mbox{16-PQAM($\Gamma$)} with~\mbox{$\Gamma=\lbrace 4,8 \rbrace$} using the PM-D. Under medium GPN regime,~\mbox{16-PQAM(4)} exceeds~\mbox{16-QAM} starting at~$18$~dB when~\mbox{16-PQAM(8)} reaches the target BER with the same~SNR value as~\mbox{16-QAM}. Athough~\mbox{16-PQAM(4)} surpasses~\mbox{16-QAM} at high~SNR, it is noted that~\mbox{16-QAM} surpasses both~\mbox{16-PQAM(4)} and~\mbox{16-PQAM(8)} in low SNR regime. This could be a benefit when considering channel coding. According to the strong GPN regime, only the~\mbox{16-PQAM(8)} reaches the target BER. 

 \begin{figure}[tb]
   \centering    
    \includegraphics[width=0.85\columnwidth]{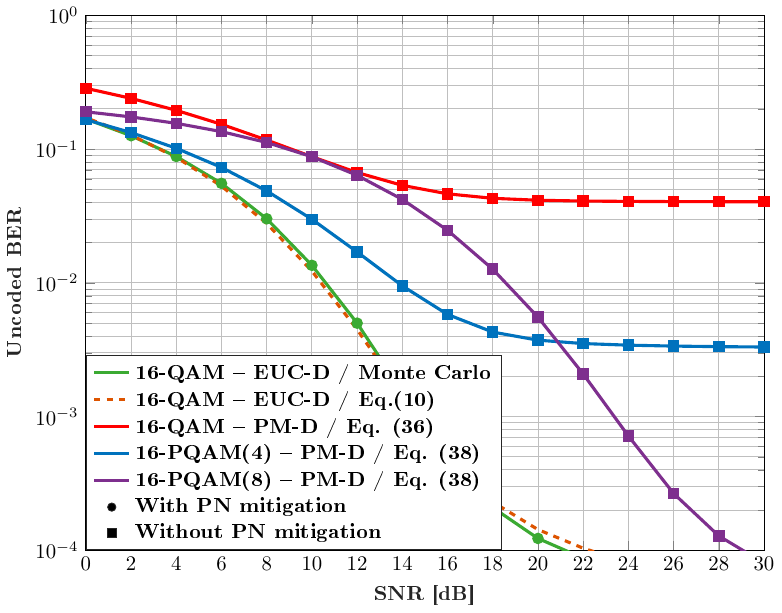}
    \caption{BER performance comparison with/without PN mitigation under strong GPN regime. We consider~$N_{\text{RF}}=N_{\text{s}}=8$, $N_{\text{pil}}=1$, $N_{\text{t}}=144$, $N_{\text{r}}=36$.}
    \label{fig:BER_PN_mitigated}
    \vspace{-0.3cm}
\end{figure}

Fig.~\ref{fig:BER_monte_carlo_vs_theoretical_PQAM}a and Fig.~\ref{fig:BER_monte_carlo_vs_theoretical_PQAM}b compare the performance obtained from Monte Carlo simulations against the one from the semi-analytical BER expression given in~\eqref{eq:ber_clo_pn_pqam} under medium and strong GPN regimes, respectively. In medium GPN regime, results return a match between the Monte Carlo and analytical curves for~\mbox{16-PQAM(8)}. For~\mbox{16-PQAM(4)}, the BER performance with~\eqref{eq:ber_clo_pn_pqam} surpasses the Monte Carlo one and thereupon can be viewed as a lower bound expression. Moreover, the high-SNR error floor caused by the strong GPN regime confirms the validity of expression~\eqref{eq:ber_clo_pn_pqam2} which only depends on the modulation order and on the GPN variance. As a consequence, whatever the channel model considered, the closed-form expressions~\eqref{eq:Pe_16_QAM_final},~\eqref{eq:ber_clo_pn_pqam2} and~\eqref{eq:ber_clo_pn_4qam}, accurately approximate the average BER expression in~high-SNR and strong GPN regimes (assuming perfect channel precoding) for~\mbox{16-QAM},~$M$-PQAM($\Gamma$) and~\mbox{4-QAM} constellations.

Fig.~\ref{fig:BER_PN_mitigated} presents a comparison between the performance obtained from: (i) semi-analytical BER expression~\eqref{eq:BER_16_QAM} for~\mbox{16-QAM} and~\eqref{eq:ber_clo_pn_pqam} for~\mbox{16-PQAM(4)} and~\mbox{16-PQAM(8)}, both associated with~\mbox{PM-D} (no PN compensation), and (ii) performance after~PN mitigation assuming~\mbox{16-QAM} with~\mbox{EUC-D}. We also illustrate the comparison between Monte Carlo after PN compensation and the semi-analytical expression~\eqref{eq:ber_nopn} when~PN does not exist~(negligible). We consider the use of a single pilot (corresponding to a pilot density of 12.5\%) for PN tracking. It can be observed that the performance with PN cancellation is better than that of~\mbox{16-PQAM} without PN cancellation. Moreover, the neutralized-PN performance corresponds to the semi-analytical BER expression~\eqref{eq:ber_nopn}.

\section{Discussions}\label{sec:discussions}

\subsection{Energy Efficiency and Computational Complexity}
Considering RF power amplifiers, which are closely linked to the transmitters’ energy consumption, the peak-to-average power ratio (PAPR) constitutes a crucial performance metric for communication systems. As demonstrated in~\cite{Bicaïs}, the PAPR of the PQAM is a strictly increasing function of the number of amplitude levels~$\Gamma$. This represents a drawback from an energy efficiency~(EE) viewpoint. 

For numerical simulations, the FC scheme was chosen to maximize the achievable SE. However, from an EE perspective, the FC scheme is not ideal when considering a high~$N_{\text{RF}}$ value~(still considering~$N_{\text{RF}} \ll N_{\text{t}}$)~\cite{Yu}. The PC scheme with or without switches was proposed as a compromise between SE and EE~\cite{Méndez-Rial,Ren,Qi,Ma}. The authors in~\cite{Yu} showed that increasing $N_{\text{RF}}$ rapidly degrades the EE in the FC architecture, since increasing the number of RF chains means increasing the number of PSs for analog precoding. As a result, the total power demand is expected to grow significantly. This is unlike the PC case, where a few number of PSs is connected to a group of RF chains and thereupon an~EE improvement. However, for a small $N_{\text{RF}}$ value such as $ N_{\text{s}}\leq N_{\text{RF}} \leq 2N_{\text{s}}-1$, as considered in our work, both EE and SE are better in FC than in PC. In short, the number of RF chains is decisive for choosing between FC or PC architecture for achieving a good EE.

In this paper, we have worked with SVD precoding which maximizes the SNR. However, this technique adds complexity at the Rx side~\footnote{Since the SVD requires a decoding step at the~Rx in contrast to ZF or MMSE which only require a precoding step at the Tx.} unlike other precoding techniques such as~ZF or minimum mean square error~(MMSE). Thus, applying ZF or MMSE could reduce the complexity compared to the SVD. As a result, it is possible to deduce the analytical expressions presented in this work. The equivalent channel matrix becomes~$\mathbf{H}_{\text{eq}}=\mathbf{H}\mathbf{F}_{\text{RF}}$. So, if we consider the ZF or MMSE precoding algorithm, we compute the digital precoding matrix~$\mathbf{F}_{\text{BB}}$ as
\begin{equation}
    \mathbf{F}^{\text{ZF}}_{\text{BB}} = \mathbf{H}_{\text{eq}}^\dagger \quad \text{and} \quad \mathbf{F}^{\text{MMSE}}_{\text{BB}} = \left(\mathbf{H}^H_{\text{eq}}\mathbf{H}_{\text{eq}} + \sigma^2\mathbf{I}_{N_\text{r}} \right)^{-1} \mathbf{H}^H_{\text{eq}},
\end{equation}
where $\left(\cdot \right)^\dagger$ denotes the Moore-Penrose pseudo-inverse. Nonetheless, the number of RF chains must be equal to the number of receive antennas and thus representing a drawback in an EE viewpoint if we assume a large number of receive antennas or a large number of users in multi-user MIMO systems.

\subsection{Perspectives}

The presence of strong GPN would limit the system's operation to low modulation orders such as \mbox{4-QAM}. This would require the use of very wide signal bands to achieve high data rates. Nevertheless, increasing signal bandwidth will add constraints on ADC/DAC by requiring very fast sampling period. Thanks to the generalized spatial modulation~(GSM) which offers a counterpart for hardware impairments~\cite{Saad}. This multi-antenna modulation technique consists in mapping some information bits on few activated antennas while using low modulation order to transmit the rest of information bits through the RF chains~\cite{Wang}. 

The GSM showed its importance in full digital MIMO by providing an alternative for achieving high throughput despite the constraint of low modulation order in the presence of strong GPN~\cite{Saad2}. Additionally, GSM has already been investigated in HP MIMO system without PN impairments in terms of BER and EE~\cite{Li2}. A perspective to this work is to study the implementation of GSM in HP MIMO system impaired by~GPN since there is no contribution proposed in the state-of-the-art to the best of our knowledge. Also, implementing MC waveforms such as OFDM and DFT-s-OFDM might be interesting to evaluate the possibility of reusing all 3GPP concepts on framing, definition of PHY signals, channel estimation and so on.

Furthermore, a new stacked intelligent metasurface~(SIM)-based concept is proposed to achieve optimal EE performance. The authors in~\cite{Jiancheng} propose a promising transceiver architecture based on multi-layer~SIM to realize precoding and combining in the wave domain. Basically, the digital and/or analog precoding stages are done in the wave domain to provide a very low hardware cost and therefore, a very high EE compared to exisiting MIMO technologies. However, according to the sub-THz properties, some constraints such as high-resolution ADC/DAC and the generated PN will certainly affect its performance. Thus, another perspective of this work is to study the performance of SIM-based MIMO systems with GPN. Furthermore, it could also be interesting to present a performance comparison with the HP MIMO system.

\section{Conclusions} \label{sec:conclusion}
We investigated the GPN impairment on SVD-based HP SU-MIMO systems for future sub-THz applications. We derived a theoretical lower bound expression on the achievable SE and closed-form~BER expressions for~\mbox{4-QAM},~\mbox{16-QAM} and $M$-PQAM($\Gamma$) under CLO architecture. We validated the theoretical expressions via Monte Carlo simulations. Moreover, we showed that the PM detector is not necessary for low modulation order such as~\mbox{4-QAM} unlike the~\mbox{16-QAM} where the system is more affected by~GPN due to the noise sensitivity when increasing the modulation order. Additionally, we highlighted that the PQAM with~$\Gamma = M/2$ is more robust to GPN compared to the~\mbox{QAM}. However, the~QAM with the PM detector could be more adequate when channel coding is considered and regarding the~$\Gamma$ increasing PAPR of the PQAM constellation. We also outlined the possibility of enhancing the system performance by alleviating the GPN with a single pilot. Finally, the closed-form BER expressions derived in this paper considering high-SNR and strong GPN regimes remain valid for all HP MIMO systems—whether SU or multi-user—irrespective of the precoding algorithm employed, provided a CLO architecture and perfect CSI are assumed.

\section*{Acknowledgment} 
Part of this work was funded by the French National Research Agency (22-PEFT-0006) as part of France 2030 and the NF-SYSTERA project.
\section*{Appendix}\label{sec:apdx}
In this appendix, we state the proof of the equations~\eqref{eq:the_se_pn_clo2},~\eqref{eq:Pe_16_QAM} and~\eqref{eq:ber_clo_pn_pqam}. 

\subsection*{A. Lower Bound Expression of the Achievable Rate} 
We have from~\eqref{eq:the_se_pn_clo}
\begin{equation}
        \text{R}_{\text{PN}} = \sum_{k=1}^{N_{\text{s}}} \log_2 \left( 1 + \frac{\rho^2 \left\lvert \mathbb{E} \left\lbrace e^{j\psi}\mathbf{u}_k^{\text{BB}^H} \mathbf{H}_{\text{eq}}  \mathbf{f}_k^{\text{BB}} \right \rbrace \right\rvert^2 }{\rho^2 \kappa_k + \mathbb{E} \left\lbrace \left \lvert e^{j\varphi}\mathbf{u}_k^{\text{BB}^H} \mathbf{W}^H_{\text{RF}} \mathbf{n} \right \rvert^2 \right\rbrace} \right),
    \label{eq:apdxthe_se_pn_clo}
\end{equation}
where~$\kappa_k = \mathbb{E} \left\lbrace \left \lvert \mathcal{M}_k s_k \right \rvert^2 \right\rbrace$ with~$\mathcal{M}_k$ is defined in~\eqref{eq:rcvd_hybp_pn_clo2_self_interf}. By considering the SVD of~$\mathbf{H}_{\text{eq}}=\mathbf{U}_{\text{BB}} \mathbf{V}\mathbf{F}^H_{\text{BB}}$, the term~$\mathbb{E} \left\lbrace e^{j\psi}\mathbf{u}_k^{\text{BB}^H} \mathbf{H}_{\text{eq}}  \mathbf{f}_k^{\text{BB}} \right \rbrace$ is simplified as follows
\begin{equation}
   \begin{aligned}
        \mathbb{E} \left\lbrace e^{j\psi}\mathbf{u}_k^{\text{BB}^H} \mathbf{H}_{\text{eq}}  \mathbf{f}_k^{\text{BB}} \right \rbrace &= \mathbb{E} \left\lbrace e^{j\psi} \right \rbrace \mathbf{u}_k^{\text{BB}^H} \mathbf{H}_{\text{eq}}  \mathbf{f}_k^{\text{BB}} \\
        &= e^{-\frac{\sigma^2_\psi}{2}} \mathbf{V}_{k,k},
    \end{aligned}
    \label{eq:apdxthe_numerator_pn2}
\end{equation}
where~$\mathbb{E} \left\lbrace e^{j\psi} \right \rbrace = e^{-\frac{\sigma^2_\psi}{2}}$ represents the  characteristic function of~$e^{j\psi}$ and~$\mathbf{V}_{k,k}=\mathbf{u}_k^{\text{BB}^H} \mathbf{H}_{\text{eq}} \mathbf{f}_k^{\text{BB}}$. The term~$\kappa_k = \mathbb{E} \left\lbrace \left \lvert \mathcal{M}_k s_k \right \rvert^2 \right\rbrace$ is given by
\begin{equation}
     \begin{aligned}
        \kappa_k &= \mathbb{E} \left\lbrace \left \lvert \mathcal{M}_k s_k \right \rvert^2 \right\rbrace \\
        &=  \mathbb{E} \left\lbrace \left\lvert e^{j\psi}\mathbf{u}_k^{\text{BB}^H} \mathbf{H}_{\text{eq}} \mathbf{f}_k^{\text{BB}} \right \rvert^2 \right \rbrace - \left\lvert \mathbb{E} \left\lbrace e^{j\psi}\mathbf{u}_k^{\text{BB}^H} \mathbf{H}_{\text{eq}}  \mathbf{f}_k^{\text{BB}} \right \rbrace \right \rvert^2 \\
        &= \mathbb{E} \left\lbrace \left\lvert \mathbf{V}_{k,k} \right \rvert^2 \right\rbrace - e^{-\sigma^2_\psi} \left\lvert \mathbf{V}_{k,k} \right \rvert^2 \\
        &= \left(1 - e^{-\sigma^2_\psi} \right) \left\lvert \mathbf{V}_{k,k} \right \rvert^2 \cdot
      \end{aligned}
    \label{eq:apdxrcvd_hybp_pn_clo2_self_interf}
\end{equation}
The last term~$\mathbb{E} \left\lbrace \left \lvert e^{j\varphi}\mathbf{u}_k^{\text{BB}^H} \mathbf{W}^H_{\text{RF}} \mathbf{n} \right \rvert^2 \right\rbrace$ can be simplified as follows
\begin{equation}
     \begin{aligned}
        \mathbb{E} \left\lbrace \left \lvert e^{j\varphi}\mathbf{u}_k^{\text{BB}^H} \mathbf{W}^H_{\text{RF}} \mathbf{n} \right \rvert^2 \right\rbrace &= \sigma^2 \mathbf{u}_k^{\text{BB}^H} \mathbf{W}^H_{\text{RF}} \mathbf{W}_{\text{RF}} \mathbf{u}_k^{\text{BB}}\\
        &= \sigma^2 \xi_k \cdot
    \end{aligned}
    \label{eq:apdxthe_noise2}
\end{equation}
By replacing~\eqref{eq:apdxthe_numerator_pn2}, \eqref{eq:apdxrcvd_hybp_pn_clo2_self_interf} and~\eqref{eq:apdxthe_noise2} in~\eqref{eq:apdxthe_se_pn_clo}, the final expression of the achievable rate when considering the CLO scheme is given by
\begin{equation}
    \begin{aligned}
        \text{R}_{\text{PN}} &= \sum_{k=1}^{N_{\text{s}}} \log_2 \left( 1 + \frac{\rho^2 e^{-\sigma^2_\psi} \left\lvert \mathbf{V}_{k,k} \right\rvert^2 }{\rho^2 \left(1 - e^{-\sigma^2_\psi}\right) \lvert \mathbf{V}_{k,k} \rvert^2 + \sigma^2 \xi_k } \right) \\
        &= \sum_{k=1}^{N_{\text{s}}} \log_2 \left( 1 + \frac{\rho^2 \left\lvert \mathbf{V}_{k,k} \right\rvert^2 }{\rho^2 \left(e^{\sigma^2_\psi} - 1\right) \lvert \mathbf{V}_{k,k} \rvert^2 + \sigma^2 e^{\sigma^2_\psi} \xi_k } \right) \cdot
    \end{aligned}        
    \label{eq:apdxthe_se_pn_clo2}
\end{equation}

\subsection*{B. Theoretical BER expression of 16-QAM constellation under GPN impairment}

The symbols on each amplitude level (red, blue or purple) as depicted in~Fig.~\ref{fig:16_QAM_representation}b, have the same detection error probability regarding the misestimation of the amplitude in each amplitude level. 

\subsubsection*{B.1. Detection Error Probability on the First and Third Amplitude Levels (red solid line and purple dashdotted line)}
\smallskip
$\\$
The symbols on the first and third amplitude level also have the same detection error probability regarding the misestimation of the phase.
From Fig.~\ref{fig:16_QAM_representation}b, we can approximate the detection error probability of the black (bk) symbol in the first amplitude level (red solid line) by
\begin{equation}
    \begin{aligned}    
      P_{e_{bk}}^{(r)} & \simeq 2 \mathcal{Q}\left(\frac{\delta_{\rho_1}}{2\sigma_{n_k}}\right) + 2 \mathcal{Q}\left(\frac{\delta_{\theta_1}}{2\sqrt{ \sigma^2_\psi  + \sigma_{n_k}^2/E_s}}\right),
    \end{aligned}
\end{equation}
where the first term and the second term represent the probability to have an error on the amplitude and on the phase, respectively. Since we have the same~$\delta_{\theta_1}$ between the symbols, we deduce the detection error probability of the four symbols on the first amplitude level as follows
\begin{equation}
    \begin{aligned}    
      P_e^{(r)} & \simeq 4 P_{e_{bk}}^{(r)} = 8 \mathcal{Q}\left(\frac{\delta_{\rho_1}}{2\sigma_{n_k}}\right) + 8 \mathcal{Q}\left(\frac{\delta_{\theta_1}}{2\sqrt{ \sigma^2_\psi  + \sigma_{n_k}^2/E_s}}\right) \cdot
    \end{aligned}
\end{equation}
Similarly to the first amplitude level, we can approximate the detection error probability of the black (bk) symbol on the third amplitude level~(purple dashdotted line) by
\begin{equation}
    \begin{aligned}    
      P_{e_{bk}}^{(p)} & \simeq 2 \mathcal{Q}\left(\frac{\delta_{\rho_2}}{2\sigma_{n_k}}\right) + 2 \mathcal{Q}\left(\frac{\delta_{\theta_4}}{2\sqrt{ \sigma^2_\psi  + \sigma_{n_k}^2/E_s}}\right) \cdot
    \end{aligned}
\end{equation}
Given that we have the same~$\delta_{\theta_4}$ between the symbols, we deduce the detection error probability of the four symbols on the first amplitude level as follows
\begin{equation}
    \begin{aligned}    
      P_e^{(p)} & \simeq 4 P_{e_{bk}}^{(p)} = 8 \mathcal{Q}\left(\frac{\delta_{\rho_2}}{2\sigma_{n_k}}\right) + 8 \mathcal{Q}\left(\frac{\delta_{\theta_4}}{2\sqrt{ \sigma^2_\psi + \sigma_{n_k}^2/E_s}}\right) \cdot
    \end{aligned}
\end{equation}
Because~$\delta_{\theta_1}=\delta_{\theta_4}$ for the $16$-QAM, the detection error probability of the symbols at the red and purple levels can be expressed by
\begin{equation}
    \begin{aligned}    
      P_e^{\left(r,p\right)} &\simeq P_e^{\left(r\right)} + P_e^{\left(p\right)}= 8 \left( \mathcal{Q}\left(\frac{\delta_{\rho_1}}{2\sigma_{n_k}}\right) + \mathcal{Q}\left(\frac{\delta_{\rho_2}}{2\sigma_{n_k}}\right) \right) \\
      &+ 16 \mathcal{Q}\left(\frac{\delta_{\theta_1}}{2\sqrt{ \sigma^2_\psi + \sigma_{n_k}^2}}\right) \cdot
    \end{aligned}
    \label{eq:apdx_Pe1}
\end{equation}

\subsubsection*{B.2. Detection Error Probability on the second Amplitude Level (blue dashed line)}
$\\ \\$
In the second amplitude level, all the eight symbols have the same detection error probability according the amplitude estimation but not for phase estimation. Thanks to the symmetry of symbols, we will derive the detection error of the colored symbols (black, red, pink and blue) and multiply each of them by 2 to obtain the total detection error probability of the symbols in the second amplitude level. Thus, we can approximate the detection error probability of the black~(bk), red~(rd), pink(pk) and blue~(be) by
\begin{equation}
    \begin{aligned} 
\left\{
    \begin{array}{ll}
        P_{e_{bk}}^{\left(b\right)} \simeq 2 \mathcal{Q}\left(\frac{\min(\delta_{\rho_1},\delta_{\rho_2})}{2\sigma_{n_k}}\right) + 2 \mathcal{Q}\left(\frac{\min(\delta_{\theta_2},\delta_{\theta_3})}{2\sqrt{ \sigma^2_\psi  + \sigma_{n_k}^2/E_s}}\right)  \\

        P_{e_{rd}}^{\left(b\right)} \simeq  2 \mathcal{Q}\left(\frac{\min(\delta_{\rho_1},\delta_{\rho_2})}{2\sigma_{n_k}}\right) + 2 \mathcal{Q}\left(\frac{\min(\delta_{\theta_2},\delta_{\theta_3})}{2\sqrt{ \sigma^2_\psi  + \sigma_{n_k}^2/E_s}}\right) \\

        P_{e_{pk}}^{\left(b\right)} \simeq 2 \mathcal{Q}\left(\frac{\min(\delta_{\rho_1},\delta_{\rho_2})}{2\sigma_{n_k}}\right) + 2 \mathcal{Q}\left(\frac{\min(\delta_{\theta_2},\delta_{\theta_3})}{2\sqrt{ \sigma^2_\psi  + \sigma_{n_k}^2/E_s}}\right) \\

       P_{e_{be}}^{\left(b\right)} \simeq 2 \mathcal{Q}\left(\frac{\min(\delta_{\rho_1},\delta_{\rho_2})}{2\sigma_{n_k}}\right) + 2 \mathcal{Q}\left(\frac{\min(\delta_{\theta_2},\delta_{\theta_3})}{2\sqrt{ \sigma^2_\psi  + \sigma_{n_k}^2/E_s}}\right) \cdot
       
    \end{array}
\right.
    \end{aligned}
\end{equation} 
 Thereupon, by considering the relation~\eqref{eq:delta_rho_theta} and assuming~$16$-QAM normalized symbols, the detection error probability of the symbols at the blue amplitude level can be expressed by
\begin{equation}
    \begin{aligned}    
      P_e^{\left(b\right)} &\simeq 2 \left( P_{e_{bk}}^{\left(b\right)} + P_{e_{rd}}^{\left(b\right)} + P_{e_{pk}}^{\left(b\right)} + P_{e_{be}}^{\left(b\right)} \right) \\
      &\simeq 2 \left( 8 \mathcal{Q}\left(\frac{\delta_{\rho_2}}{2\sigma_{n_k}}\right) + 8 \mathcal{Q}\left(\frac{\delta_{\theta_2}}{2\sqrt{ \sigma^2_\psi  + \sigma_{n_k}^2}} \right) \right) \\
      &\simeq 16 \left(\mathcal{Q}\left(\frac{\delta_{\rho_2}}{2\sigma_{n_k}} \right) + \mathcal{Q}\left(\frac{\delta_{\theta_2}}{2\sqrt{ \sigma^2_\psi  + \sigma_{n_k}^2}} \right) \right) \cdot
    \end{aligned}
    \label{eq:apdx_Pe2}
\end{equation}
Therefore, we can deduce the detection error probability of the~$16$-QAM at the $k^{th}$ stream as follows
\begin{equation}
    \begin{aligned}
        P_{e_k} &= \frac{1}{M} \sum_{s \in \mathcal{C}} P\left(\hat{s}_k \neq s_k | s\right) \simeq \frac{1}{16} \left( P_e^{\left(r,p\right)} + P_e^{\left(b\right)}\right)\\
        & \simeq \frac{1}{2} \left(\mathcal{Q}\left(\frac{\delta_{\rho_1}}{2\sigma_{n_k}}\right) + 3 \mathcal{Q}\left(\frac{\delta_{\rho_2}}{2\sigma_{n_k}}\right)\right) \\ 
      &+ \mathcal{Q}\left(\frac{\delta_{\theta_1}}{2\sqrt{ \sigma^2_\psi  + \sigma_{n_k}^2}}\right) + \mathcal{Q}\left(\frac{\delta_{\theta_2}}{2\sqrt{ \sigma^2_\psi  + \sigma_{n_k}^2}}\right) \cdot
    \end{aligned}
    \label{eq:apdx_Pe_total}
\end{equation}

\subsection*{C. Theoretical BER expression of $M$-ary PQAM($\Gamma$) constellation under GPN impairment}
The authors in~\cite{Bicaïs} presents the BER expression of the~$M$-PQAM($\Gamma$) for a SC system impaired by strong GPN considering an AWGN channel by
\begin{equation}
 \label{eq:apdx_ber_clo_pn_pqam}
    \begin{split}
        P_{be} & \simeq \frac{2}{\log_2(M)} \Biggl( \mathcal{Q}\left(\sqrt{\frac{6 \cdot E_s / N_0}{\left(4\Gamma^2-1\right)} } \right)  \\
        &+ \mathcal{Q}\left( \frac{\pi \Gamma }{M\sqrt{\sigma^2_\psi + \frac{N_0}{2\cdot E_s}}} \right) \Biggr),
    \end{split}
\end{equation}
where~$N_0 = 2\sigma^2_n$ is the noise power spectral density. By inserting~\eqref{eq:sigma_thermal_noise} into~\eqref{eq:sigma2_n} and by using the new expression of~\eqref{eq:sigma2_n} in~\eqref{eq:apdx_ber_clo_pn_pqam}, we can approximate the BER expression of the~$M$-PQAM($\Gamma$) (assuming normalized PQAM symbols, \textit{i.e.},~$E_s = 1$ and $N_0 = 2\sigma^2_{n_k}$) at the~$k^{th}$ stream as follows
\begin{equation}
 \label{eq:apdx_ber_clo_pn_pqam2}
    \begin{split}
        P_{e_k} & \simeq \frac{2}{\log_2(M)} \Biggl( \mathcal{Q}\left(\sqrt{\frac{6 \left \lvert \mathbf{V}_{k,k} \right\rvert^2 \digamma }{\left(4\Gamma^2-1\right)\omega \left \lvert \xi_k\right\rvert} } \right)  \\
        &+ \mathcal{Q}\left( \frac{\pi \Gamma }{M\sqrt{\sigma^2_\psi + \frac{\omega \left \lvert \xi_k\right\rvert}{2\left \lvert \mathbf{V}_{k,k} \right\rvert^2  \digamma} }} \right) \Biggr) \cdot
    \end{split}
\end{equation}

\bibliographystyle{IEEEtran}
\bibliography{IEEEabrv, references.bib}
\end{document}